\documentclass[a4paper,fleqn,usenatbib]{mnras}

\usepackage{newtxtext,newtxmath}

\usepackage[T1]{fontenc}
\usepackage{ae,aecompl}

\usepackage{graphicx}	
\usepackage{amsmath}	
\usepackage{amssymb}	
\usepackage{verbatim}
\usepackage{tipa}

\title[SALT Spectra of \textit{WISE}-selected Obscured Quasars]{Characterizing the \textit{WISE}-selected Heavily Obscured Quasar Population with Optical Spectroscopy from the Southern African Large Telescope}

\author[Hviding et al.]{Raphael E. Hviding,$^{1}$\thanks{E-mail: Raphael.E.Hviding.18@Dartmouth.edu}
Ryan C. Hickox,$^{1}$
Kevin N. Hainline,$^{2}$
\newauthor Christopher M. Carroll,$^{1}$
Michael A. DiPompeo,$^{1}$
Wei Yan,$^{1}$
\newauthor and Mackenzie L. Jones$^{1}$
\\
$^{1}$Department of Physics and Astronomy, Dartmouth College, Hanover, NH 03755, USA\\
$^{2}$Steward Observatory, University of Arizona, Tucson, AZ 85721, USA\\
}

\date{Accepted XXX. Received YYY; in original form ZZZ}

\pubyear{2017}

\begin{document}
\label{firstpage}
\pagerange{\pageref{firstpage}--\pageref{lastpage}}
\maketitle

\begin{abstract}
We present the results of an optical spectroscopic survey of 46 heavily obscured quasar candidates. Objects are selected using their mid-infrared (mid-IR) colours and magnitudes from the Wide-Field Infrared Survey Explorer (\textit{WISE}) and their optical magnitudes from the Sloan Digital Sky Survey (SDSS). Candidate Active Galactic Nuclei (AGNs) are selected to have mid-IR colours indicative of quasar activity and lie in a region of mid-IR colour space outside previously published X-ray based selection regions. We obtain optical spectra for our sample using the Robert Stobie Spectrograph on the Southern African Large Telescope. Thirty objects (65\%) have identifiable emission lines, allowing for the determination of spectroscopic redshifts. Other than one object at $z\sim2.6$, candidates have moderate redshifts ranging from $z=0.1$ to $0.8$ with a median of 0.3. Twenty-one (70\%) of our objects with identified redshift (46\% of the whole sample) are identified as AGNs through common optical diagnostics. We model the spectral energy distributions of our sample and found that all require a strong AGN component, with an average intrinsic AGN fraction at 8$\,\mu$m of 0.91. Additionally, the fits require large extinction coefficients with an average $E(B-V)_\textrm{AGN} = 17.8$ (average $A(V)_\textrm{AGN} = 53.4$). By focusing on the area outside traditional mid-IR photometric cuts, we are able to capture and characterise a population of deeply buried quasars that were previously unattainable through X-ray surveys alone. 
\end{abstract}



\section{Introduction}
\begin{table*}
\begin{center}
  \caption{Sample Photometric Properties}
  \label{tab:photodata}
\setlength{\tabcolsep}{3.6pt}
  \begin{tabular}{ccccccccccrccc} 
  \hline \hline SDSS Name & Obs. Date & Obs. Time$\,^\text{a}$ & \multicolumn{5}{c}{SDSS Photometry$\,^\text{b}$} & \multicolumn{4}{c}{\textit{WISE} Photometry$\,^\text{c}$} & \multicolumn{2}{c}{\textit{WISE} Colours} \\
  & (y/m/d) & (s) & $u$ & $g$ & $r$ & $i$ & $z$ & \textit{W1} & \textit{W2} & \textit{W3} & \textit{W4} & \textit{W1}$-$\textit{W2} & \textit{W2}$-$\textit{W3} \\
\hline 
J000224.02$-$014215.6 & 2015/06/14 & 2400 & 22.94 & 21.70 & 20.17 & 19.56 & 19.40 & 15.37 & 14.39 & 10.20 & 6.93 & 0.98 & 4.19\\
J000956.62$-$002713.5 & 2015/06/13 & 2336 & 22.44 & 21.58 & 20.15 & 19.55 & 19.17 & 15.99 & 15.11 & 11.10 & 6.75 & 0.88 & 4.01\\
J002628.56$+$011734.1 & 2015/07/15 & 2400 & 21.75 & 20.68 & 19.28 & 18.60 & 18.20 & 14.82 & 13.76 & 9.65 & 6.54 & 1.06 & 4.11\\
J005503.68$-$015753.3 & 2015/08/08 & 2400 & 21.99 & 20.02 & 18.78 & 18.42 & 18.18 & 15.16 & 14.35 & 10.39 & 6.98 & 0.81 & 3.96\\
J005621.72$+$003235.7 & 2015/07/27 & 2400 & 22.42 & 21.47 & 20.70 & 19.68 & 19.64 & 15.40 & 14.58 & 9.74 & 6.53 & 0.82 & 4.84\\
J014737.42$+$014955.7 & 2014/11/29 & 2400 & 21.23 & 20.12 & 19.06 & 18.73 & 18.34 & 14.56 & 13.68 & 10.16 & 6.93 & 0.88 & 3.52\\
J025759.11$-$004136.9 & 2014/12/27 & 2400 & 22.89 & 21.01 & 19.86 & 19.30 & 18.94 & 15.61 & 14.49 & 10.05 & 6.67 & 1.12 & 4.44\\
J035524.81$-$001524.7 & 2015/10/15 & 2400 & 21.70 & 20.76 & 19.60 & 20.05 & 18.89 & 15.82 & 14.86 & 10.50 & 6.98 & 0.96 & 4.36\\
J045401.15$-$003822.2 & 2014/12/30 & 2400 & 22.02 & 20.73 & 19.82 & 19.18 & 19.00 & 15.26 & 14.16 & 9.89 & 6.88 & 1.10 & 4.27\\
J051133.90$+$002819.0 & 2015/09/22 & 2400 & 21.93 & 20.06 & 18.73 & 18.44 & 18.03 & 14.23 & 13.42 & 9.67 & 6.50 & 0.81 & 3.75\\
J051135.47$-$012826.7 & 2015/01/12 & 2211 & 22.43 & 20.71 & 19.51 & 18.78 & 18.03 & 14.12 & 13.11 & 9.04 & 6.60 & 1.01 & 4.07\\
J052016.18$-$013504.1 & 2014/11/29 & 2400 & 21.11 & 20.13 & 18.96 & 18.27 & 18.04 & 14.84 & 13.83 & 9.61 & 6.98 & 1.01 & 4.22\\
J053056.03$-$010012.5 & 2015/10/16 & 2400 & 24.00 & 21.72 & 20.67 & 20.11 & 19.50 & 14.80 & 14.04 & 9.41 & 6.69 & 0.76 & 4.63\\
J053241.28$-$002936.4 & 2014/12/11 & 2400 & 21.35 & 20.38 & 19.21 & 18.71 & 18.18 & 14.79 & 13.98 & 9.76 & 6.98 & 0.81 & 4.22\\
J061340.30$-$005119.8 & 2014/12/27 & 2400 & 24.41 & 21.12 & 19.66 & 18.85 & 18.29 & 14.31 & 13.52 & 9.30 & 6.83 & 0.79 & 4.22\\
J063354.53$+$001640.6 & 2015/01/12 & 2400 & 23.65 & 21.11 & 19.54 & 18.17 & 17.35 & 13.75 & 12.87 & 9.34 & 6.80 & 0.88 & 3.53\\
J081331.40$-$000630.6 & 2015/01/09 & 2400 & 22.13 & 20.11 & 18.93 & 18.52 & 18.30 & 14.43 & 13.59 & 9.28 & 6.56 & 0.84 & 4.31\\
J081733.09$-$011248.7 & 2015/01/11 & 2400 & 21.12 & 20.30 & 19.54 & 18.96 & 18.69 & 14.34 & 13.58 & 9.90 & 7.00 & 0.76 & 3.68\\
J083448.48$+$015921.1 & 2015/01/22 & 2400 & 22.76 & 20.64 & 21.20 & 20.99 & 19.79 & 17.09 & 15.50 & 9.62 & 6.88 & 1.59 & 5.88\\
J100817.25$-$005731.5 & 2015/01/21 & 2250 & 21.85 & 20.93 & 20.07 & 19.29 & 18.85 & 15.85 & 14.78 & 10.08 & 6.88 & 1.07 & 4.70\\
J100848.15$+$011801.4 & 2015/01/11 & 2400 & 23.42 & 21.83 & 20.71 & 20.33 & 20.29 & 16.70 & 15.49 & 9.76 & 6.58 & 1.21 & 5.73\\
J113954.32$-$010500.9 & 2015/01/21 & 2400 & 21.96 & 20.42 & 19.51 & 19.02 & 18.75 & 15.64 & 14.88 & 10.28 & 6.83 & 0.76 & 4.60\\
J114353.77$+$010947.2 & 2015/05/08 & 2400 & 22.83 & 21.47 & 20.33 & 19.32 & 19.35 & 15.55 & 14.68 & 9.85 & 6.76 & 0.87 & 4.83\\
J130120.87$+$004422.6 & 2015/05/09 & 2400 & 21.45 & 21.04 & 19.99 & 18.98 & 18.99 & 15.20 & 14.29 & 10.29 & 6.99 & 0.91 & 4.00\\
J131015.79$-$010321.6 & 2015/01/22 & 2400 & 22.50 & 21.13 & 19.99 & 19.45 & 19.12 & 16.02 & 14.80 & 9.48 & 6.55 & 1.22 & 5.32\\
J135925.18$-$010451.1 & 2015/05/18 & 2400 & 20.64 & 20.05 & 19.17 & 18.67 & 18.34 & 15.10 & 14.26 & 10.26 & 6.92 & 0.84 & 4.00\\
J142104.04$+$013337.6 & 2015/03/15 & 2400 & 21.40 & 20.12 & 19.31 & 18.82 & 18.56 & 15.45 & 14.72 & 9.96 & 6.73 & 0.73 & 4.76\\
J142941.46$-$014119.6 & 2015/04/09 & 2400 & 20.97 & 20.58 & 19.98 & 19.34 & 19.01 & 15.26 & 14.43 & 9.75 & 7.00 & 0.83 & 4.68\\
J150858.17$+$004314.0 & 2015/05/13 & 2400 & 23.02 & 21.32 & 20.11 & 19.33 & 19.27 & 15.24 & 14.14 & 9.64 & 6.68 & 1.10 & 4.50\\
J154826.03$+$004615.3 & 2015/05/13 & 2400 & 22.46 & 21.60 & 20.21 & 19.50 & 19.11 & 14.79 & 13.97 & 10.13 & 6.86 & 0.82 & 3.84\\
J154909.79$+$011940.5 & 2015/05/17 & 2400 & 20.93 & 20.53 & 19.85 & 19.56 & 19.08 & 15.14 & 14.34 & 10.42 & 6.75 & 0.80 & 3.92\\
J155048.36$+$002859.7 & 2015/03/30 & 2400 & 21.20 & 20.63 & 19.78 & 19.06 & 18.93 & 15.39 & 14.45 & 10.48 & 7.00 & 0.94 & 3.97\\
J155353.96$-$001029.3 & 2015/03/15 & 2400 & 23.00 & 21.01 & 19.87 & 19.35 & 18.94 & 15.62 & 14.91 & 10.79 & 6.93 & 0.71 & 4.12\\
J162201.42$+$002931.8 & 2015/04/12 & 2400 & 22.98 & 20.80 & 19.57 & 19.04 & 18.50 & 14.82 & 13.95 & 9.66 & 6.54 & 0.87 & 4.29\\
J164434.17$+$012839.5 & 2015/03/30 & 2100 & 26.58 & 20.95 & 19.62 & 19.04 & 18.49 & 14.56 & 13.76 & 9.56 & 6.93 & 0.80 & 4.20\\
J171952.75$-$001552.8 & 2015/04/12 & 2250 & 23.08 & 20.27 & 18.94 & 18.37 & 18.10 & 14.72 & 13.88 & 9.86 & 6.85 & 0.84 & 4.02\\
J180408.11$+$010004.0 & 2015/05/13 & 2400 & 23.45 & 21.05 & 19.68 & 19.13 & 18.84 & 14.46 & 13.59 & 9.94 & 6.98 & 0.87 & 3.65\\
J195611.27$-$000718.0 & 2015/05/16 & 2400 & 21.88 & 21.01 & 20.11 & 19.47 & 19.08 & 15.05 & 14.19 & 9.91 & 6.55 & 0.86 & 4.28\\
J204839.63$+$005449.3 & 2015/05/15 & 2400 & 21.06 & 20.10 & 19.37 & 19.10 & 18.50 & 15.36 & 14.56 & 10.04 & 6.97 & 0.80 & 4.52\\
J211845.16$-$003914.6 & 2015/05/15 & 2400 & 21.77 & 20.15 & 19.33 & 18.86 & 18.59 & 15.55 & 14.80 & 10.56 & 6.72 & 0.75 & 4.24\\
J212649.41$-$000257.7 & 2015/05/19 & 2400 & 24.08 & 21.97 & 20.83 & 20.24 & 19.77 & 16.81 & 16.07 & 10.66 & 6.83 & 0.74 & 5.41\\
J221812.68$-$013442.7 & 2015/05/18 & 2400 & 21.44 & 20.35 & 19.15 & 18.69 & 18.14 & 14.77 & 14.05 & 9.96 & 6.72 & 0.72 & 4.09\\
J221817.26$+$003623.6 & 2015/05/19 & 2300 & 21.49 & 20.24 & 18.73 & 18.55 & 17.90 & 14.86 & 13.94 & 9.93 & 6.67 & 0.92 & 4.01\\
J223059.01$-$000057.5 & 2015/05/22 & 2400 & 22.32 & 20.80 & 19.58 & 18.97 & 18.62 & 14.02 & 13.22 & 9.36 & 6.66 & 0.80 & 3.86\\
J233240.87$-$011557.9 & 2015/06/12 & 2400 & 21.88 & 20.93 & 19.74 & 18.71 & 18.48 & 14.98 & 14.12 & 9.96 & 6.88 & 0.86 & 4.16\\
J234956.07$+$014110.2 & 2015/06/18 & 2400 & 23.59 & 21.21 & 20.60 & 20.21 & 19.36 & 14.82 & 13.92 & 9.91 & 6.84 & 0.90 & 4.01\\
\hline
  \multicolumn{14}{l}{$^\text{a}$\,The objects were observed with three integrations, each approximately a third of the total observation time.}\\
  \multicolumn{14}{l}{$^\text{b}$\,Taken from the SDSS DR9 catalogue. Photometry is given in AB magnitudes.}\\
  \multicolumn{14}{l}{$^\text{c}$\,Taken from the All\textit{WISE} catalogue. Photometry is given in Vega magnitudes.}\\
  \end{tabular}
  \end{center}
\end{table*}
Active galactic nuclei (AGNs), luminous galaxy cores powered by accretion onto a central supermassive black hole (SMBH), have played a prominent role in understanding galaxy evolution ever since the discovery of quasars, the most luminous AGNs, in the late 1960s \citep{schmidt63qso}. Since then, a multitude of AGN classifications have arisen to form the so-called `AGN Zoo' that has presented itself over the last few decades \citep[e.g.][]{urry95,PadovaniSub}. AGN unification supposes that the nuclear emission is obscured by varying amounts of dust along our line of sight and divides AGNs into two populations: unobscured (type I) and obscured (type II) sources.

Possible AGN fuelling mechanisms have proposed an evolutionary process by which the interaction of merging gas-rich galaxies drives material into the nuclear region \citep[e.g.][]{kauf00merge,hopk06apjs,hopk08frame1}. In the evolutionary framework, in-falling material is the source of the optical and ultraviolet obscuration, which is eventually removed through black hole feedback processes which have been proposed to affect the galaxy on larger scales \citep{fabi12feed,alex12bh}. Supposing the extinction in obscured quasars is indeed from the nuclear region, these objects provide insight as to the fuelling process and overall evolution of AGN activity. However, given the intrinsic difficulty in detecting these objects, identification of obscured AGNs is not yet entirely complete nor accurate. In order to further the understanding of galaxy evolution and quasar processes, it is desirable to characterise objects that are missed from traditional quasar surveys. 

In general, X-ray surveys are considered to be the most efficient and effective method for selecting quasars \citep[e.g.][]{bran10,alex12bh}. X-ray emission in AGN is believed to originate near the central SMBH from Comptonization of ultraviolet and optical photons from the accretion disk \citep[e.g.][]{Haardt91Xray}. Not only do X-ray surveys suffer from little quiescent galaxy contamination as stars produce little intrinsic X-ray radiation, they are further able to recover large samples of obscured AGNs due to the penetrating nature of light at these energetic wavelengths \citep[e.g.][]{Vignali14obsc}. 

AGNs are also selected using observations at mid-infrared (mid-IR) wavelengths due to the emission believed to originate from dust surrounding and heated by the accretion region \citep{sand96}. Efforts have been made to define mid-IR colour selection in order to capture objects demonstrating this mid-IR excess. For example, \citet[herafter M12]{mate12xmmwise} makes use of X-ray selected AGN in order to define mid-IR colour selection cuts. In addition, \citet[hereafter S12]{ster12wise} creates a colour selection based on modeling spectral energy distributions (SEDs) with varying levels of dust. It is important to note that these selection criteria are not identical, where the S12 photometric cuts are able to recover objects at a higher obscuration level which are overlooked by X-ray based criteria. This discrepancy between the two mid-IR selection criteria between M12 and S12 is emphasised in \citet[hereafter Y16]{Yuan16T2QSO}, where only 34 per cent of their spectroscopically selected type II quasars are recovered by the M12 selection wedge. 

Further evidence has been found for objects that are missed in X-ray AGN surveys as demonstrated by Yan et al. (in prep.) which studied the four most obscured optically identified AGN in \citet[herafter H14]{hain14salt} with follow up \textit{Nuclear Spectroscopic Telescope Array} (\textit{NuSTAR}) observations. Three out of the four observed objects were not detected with integrations ranging from $26-40\,$ks, implying the intrinsic X-ray flux is attenuated by obscuring columns as high as $N_\text{H}\sim10^{25}\text{cm}^{-2}$. The existence of these objects combined with the results from Y16 suggest that there exists a population of heavily obscured quasars that are missed in X-ray surveys but are recoverable through their mid-IR photometry.

Typically, moderate redshift ($z$ < 1) type II quasar samples are selected based on emission line properties from spectroscopic observations. \citet[][herafter Z03]{zakamskaetal2003}, \citet[][herafter R08]{reye08qso2}, and Y16 have all made use of the Sloan Digital Sky Survey (SDSS), an optical photometric and spectroscopic survey \citep{sdssIII11}, to find samples consisting of several thousand quasars. However, despite their accuracy, spectroscopic surveys are time-intensive, cover smaller fields, and cannot go as deep as their photometric counterparts. In addition, there is a lack of photometric techniques for isolating the obscured AGN population from other AGN and galaxies. It is therefore desirable to accurately identify type II quasars solely through their photometric properties.  

Selecting AGNs at mid-IR wavelengths recovers large numbers of quasars across the entire sky because of large area infrared photometric surveys. This has been only reinforced by the \textit{Wide-Field Infrared Survey Explorer} (\textit{WISE}), an all-sky photometric survey in four mid-IR wavebands, 3.4, 4.6, 12, and $22\,\mu$m (hereafter \textit{W1, W2, W3,} and \textit{W4} respectively) \citep{wrig10wise}. \textit{WISE} has allowed for the identification of order $10^5$ of quasars across the whole sky with various selection criteria \citep{Secrest15WISE}. \citet[hereafter L13]{lacy13spec} and H14 have used follow-up spectroscopy to reinforce the identification of quasars based on infrared photometry alone. 

The purpose of this work is to characterise a subset of candidate heavily obscured AGN that lie outside traditional mid-IR selection criteria based on X-ray surveys. Given the potential high levels of obscuration present in the sample, it is possible that some are undetected even in the X-rays and as such are a missing contribution to the X-ray selected AGN population. These object may then only be recoverable through mid-IR colour selection. 

We describe the criteria used in selecting our objects in Section~\ref{sec:selection}. In Sections~\ref{sec:reduction} and~\ref{sec:emission} we detail the reduction of optical SALT spectra and conduct spectral analysis including redshift determination and the measurement of emission line diagnostics. In Section~\ref{sec:SED} we model the Spectral Energy Distributions (SEDs) for objects with spectroscopic redshifts. Our spectra are then combined in Section~\ref{sec:composite} which evaluates the composite spectra in the context of the existing literature. Objects without secure redshifts due to lack of identifiable emission features are analysed in Section~\ref{sec:noemiss}. Section~\ref{sec:interesting} highlights several objects that may follow-up observation or analysis. Finally, our conclusion and discussion are presented in Section~\ref{sec:conclusion} along with potential future research. 

Throughout this work we use the terms AGN and quasar interchangeably. We make use of a standard $\Lambda$CDM cosmology with $H_0 = 71\,$km$\,$s$^{-1}\,$Mpc$^{-1}$, $\Omega_{M} = 0.27$, and $\Omega_{\Lambda} = 0.73$ \citep{komatsu2011}.

\begin{figure}
	\centering
	\includegraphics[width=\columnwidth]{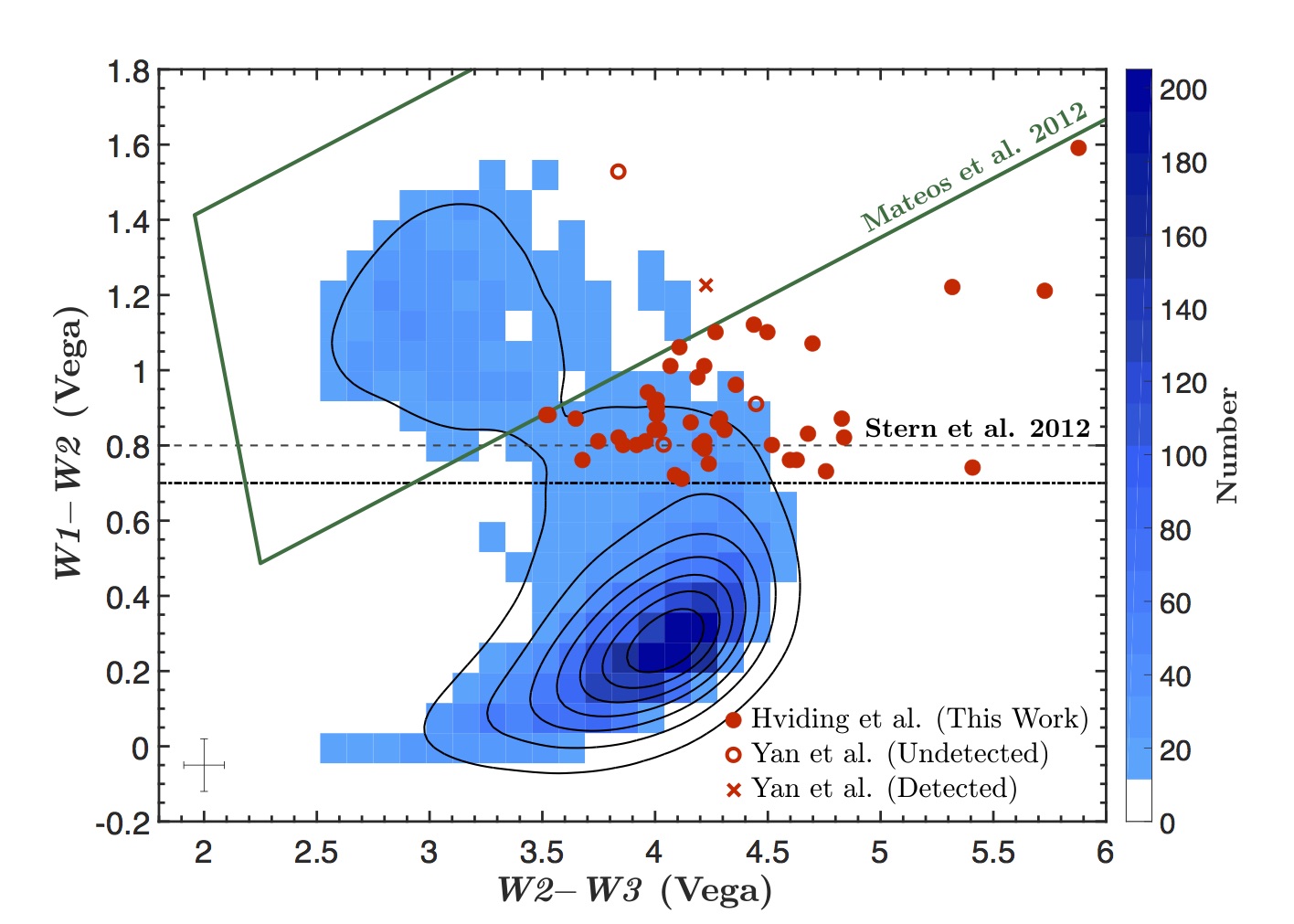}
	\caption{The \textit{WISE} colour plot with \textit{W2}$-${W3} against \textit{W1}$-$\textit{W2} with measurements in Vega magnitudes. Candidate obscured AGNs are plotted as filled red circles with typical error bars shown in the bottom left corner (this work). Open circles and crosses show \textit{NuSTAR} X-ray detected and non-detected objects respectively from Yan et al. (in prep.). Black lines show the S12 line and relaxed line. Green lines outline the M12 wedge. In blue we plot a 2D histogram with corresponding isolines of a subset of SDSS-matched \textit{WISE} sources that satisfy our \textit{W4} criteria.} 
    \label{fig:wisecolourcolour}
\end{figure}
\begin{figure*}
	\centering
	\includegraphics[width=2.09\columnwidth]{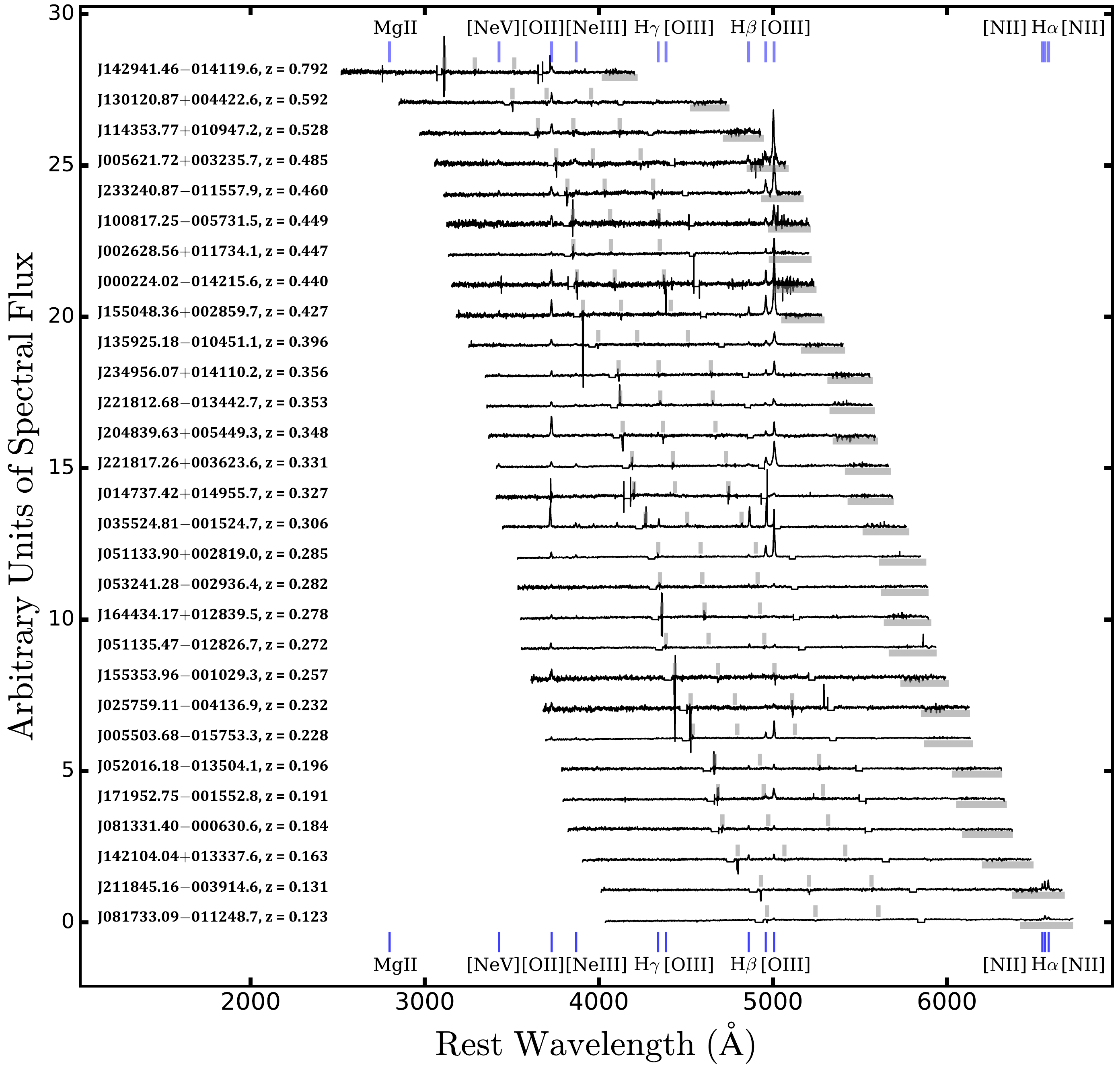}
	\caption{SALT long-slit rest frame spectroscopy of heavily obscured quasar candidates with $z<0.8$. Emission lines locations are shown with blue lines labeled at the top. Due to the nature of the stationary mirror on SALT, the objects are plotted only with relative spectral intensity, where total flux of each spectrum is normalised to unity. Telluric lines are denoted with with vertical gray bars and areas of high atmospheric disturbance with horizontal gray bars.} 
    \label{fig:G_paper}
\end{figure*}
\begin{figure*}
	\centering
	\includegraphics[width=2.09\columnwidth]{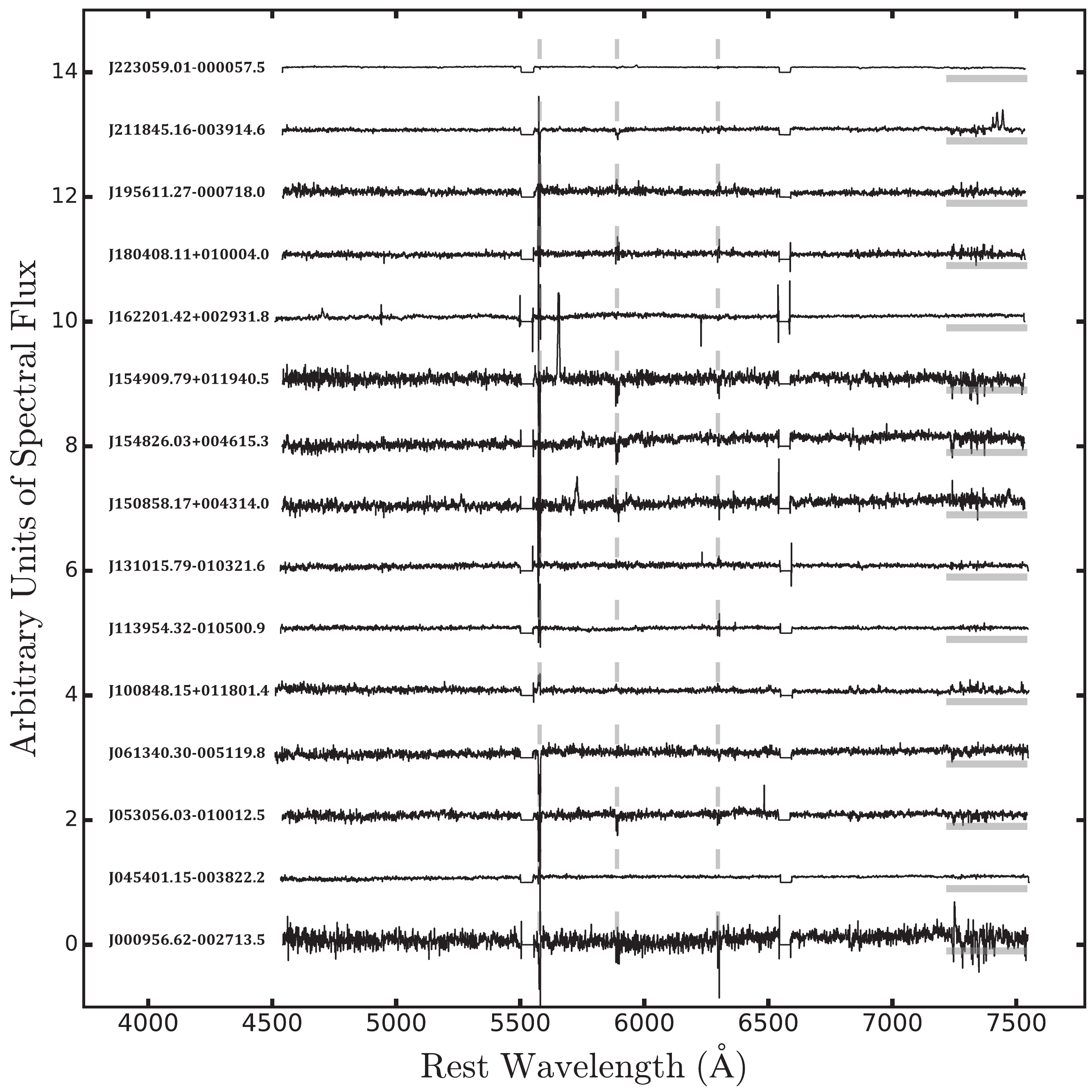}
	\caption{SALT long-slit observed spectroscopy of heavily obscured quasar candidates with no identifiable emission features. Due to the nature of the stationary mirror on SALT, the objects are plotted only with relative spectral intensity, where total flux of each spectra is normalised to unity. Telluric lines are denoted with with vertical gray bars and areas of high atmospheric disturbance with horizontal gray bars.} 
    \label{fig:No_z}
\end{figure*}

\section{Selection and Observation}
\label{sec:selection}
We base our photometric and positional criterion based on the 'Group 1' sample in H14. Our objects are selected using the same RA and Dec restrictions in order to select targets that could be observed efficiently with SALT. Furthermore, given that our observational setup was similar to that of H14, our sample was subject to an identical optical magnitude requirement of $g<22$ in order to obtain high signal to noise spectra. Optical photometric data was taken from SDSS Data Release 9 \citep[hereafter DR9]{ahn12sdss}. 

In order to find objects that are missed in mid-IR selection methods based on hard X-ray samples, our sample focuses on selecting objects outside of these regions. To reiterate, this is motivated by the Y16 sample, which found obscured AGN that do not satisfy X-ray based mid-IR selection criteria, and Yan et al. (in prep.), where X-ray undetected quasars again lie outside of these selection criteria. Here we briefly summarise the mid-IR photometric cuts for the sample in this work. Mid-IR photometric data was taken from the updated All\textit{WISE} catalogue \citep{allWISE+13}.

Our sample satisfies the same mid-IR criteria as the 'Group 1' sample from H14. Objects are selected to satisfy a relaxed S12 colour cut with \textit{W1}$-$\textit{W2} > 0.7 (opposed to 0.8). Motivated by the large AGN fraction of the Spitser $24\,\mu$m bright objects in L13, we restrict candidates to those with \textit{W4}$ \leq 7$ to capture objects with bright mid-IR emission. To select objects in a similar luminosity and redshift range we apply an lower limit in \textit{W4}, choosing objects with \textit{W4}$ \geq 6.5$. We apply an additional criterion, requiring our candidates to lie below the M12 wedge defined in equation 1 of that work in order to probe an area of \textit{WISE} colour space inhabited by objects with high extinction. We note that the $W4$ flux cut removes 97\% of objects that satisfy our \textit{WISE} colour criteria. After applying the $W4$ criterion, we apply the optical flux cut of $g < 22$, which has a relatively small effect on the sample, removing a further 8\% of the objects that satisfy the \textit{WISE} colour and W4 magnitude cuts.

Figure~\ref{fig:wisecolourcolour} plots our targets in \textit{WISE} colour space relative to selection criteria used in this work. The objects are also displayed over a density plot and contours of a subset from an SDSS-matched \textit{WISE} sample covering 3216$\,$deg$^2$ that also satisfy our W4 magnitude selection criteria. After applying the remaining selection criteria to the SDSS-matched \textit{WISE} sample we found that 660 remained, giving us an approximate sky density of one object per every 5 square degrees indicating that our objects make up a rare subset of all objects satisfying our $W4$ and $g$ band flux cuts, which have an approximate density of one per every $0.4$ square degrees.

Following the selection, the sample is observed with the Robert Stobie Spectrograph (RSS) instrument on SALT \citep{Kobulnicky03RSS,Smith06RSS}. Similar to H14, we used long-slit mode with a $1.5''$ slit and the PG0900 grating. We changed the grating angle used in H14 to 15.88$^{\circ}$ in order to include longer wavelengths, thereby increasing the likelihood of detecting [OIII]\textlambda5007$\,$\AA\ and H$\beta$\textlambda$4861\,$\AA\ emission. The final setup provided a spectral range of $4486-7534\,$\AA\ with a spectral resolution of $5.7\,$\AA\ at a central wavelength of $6041\,$\AA. We observed each object with three separate integrations, each for 800s and a total of 2400s per target. Some integrations were shortened due to atmospheric or weather interference. Table~\ref{tab:photodata} lists our 46 objects and relevant observational information from our survey, along with their photometric magnitudes and relevant colours in the available All\textit{WISE} or SDSS DR9 bands.

\section{Spectral Reduction}
\label{sec:reduction}
The observational data were reduced following the pipeline outlined in H14. The data were mosaiced, bias reduced, and dark subtracted using the SALT pipeline. The following reduction made use of standard IRAF\footnote{IRAF is distributed by the National Optical Astronomy Observatory, which is operated by the Association of Universities for Research in Astronomy (AURA), Inc., under cooperative agreement with the National Science Foundation.} scripts. We applied a wavelength solution to our objects based on Xenon and Argon arc lamp spectra. Using the three observations, bad pixel maps were generated to account for cosmic rays. The images were then background subtracted and median combined. After applying an airmass extinction correction, the objects were extracted using a $2.7''$ aperture centred on the continuum. Finally, the objects were relative flux calibrated with respect to standard star observations made available by SALT and corrected for heliocentric velocity. Due to the nature of the fixed primary mirror on SALT, for which the effective pupil size changes during the observation, it is difficult to determine absolute flux calibrations from the spectra alone. We are therefore only able to measure relative fluxes of emission lines. This is sufficient as the present study will only use flux ratios to identify quasars in our sample.

Thirty objects exhibit more than one emission feature, allowing for a visual identification of prominent lines. Figure~\ref{fig:G_paper} plots these objects, with the exception of one high redshift ($z = 2.587$) source examined in closer detail in Section~\ref{sec:interesting}, in their rest frame utilizing redshifts determined in Section~\ref{sec:emission}. One object, J063354\footnote{Shortened names are used for convenience since there is no degeneracy.}, exhibits a stellar spectrum and is therefore excluded from the following analysis. The remaining fifteen objects in the sample have a single unidentifiable emission line or no emission features and are plotted in Figure~\ref{fig:No_z}. Sections~\ref{sec:emission},~\ref{sec:SED},~and~\ref{sec:composite} shall only consider objects with identifiable redshifts from two or more lines while Section~\ref{sec:noemiss} will present analysis on objects without identified redshifts.  

\section{Emission Line Analysis}
\label{sec:emission}
\begin{figure}
	\centering
	\includegraphics[width=\columnwidth]{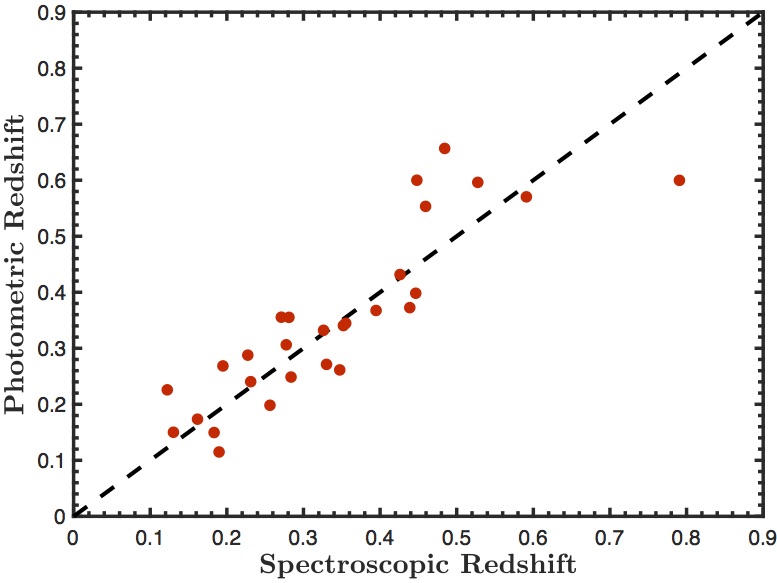}
	\caption{Measured spectroscopic redshift plotted against existing SDSS DR12 photometric redshift from \citet{Beck16drphot}. All of our objects have redshifts derived spectroscopically that are consistent with those generated from photometry alone with an $R^2$ of 0.73.} 
    \label{fig:zcomp}
\end{figure}
\begin{figure}
	\centering
	\includegraphics[width=\columnwidth]{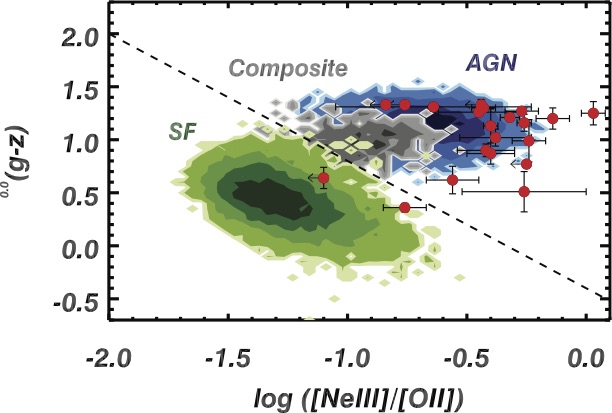}
	\caption{Colour-excitation plot for objects with [OII] and [NeIII] measurements or [NeIII] upper limits, using the \citet{trou11optx} diagnostic.  For reference, we also plot SDSS galaxies by classification using the BPT diagnostic; contours designate galaxies as star-forming (green), composite (gray), and AGN (blue).} 
    \label{fig:TBT_contour}
\end{figure}
In order to determine redshifts for our sample, we estimate preliminary redshifts from the identified lines in Section~\ref{sec:reduction}. We search for the following lines by predicting their observed frame wavelengths for each object: MgII\textlambda$2798\,$\AA, [NeV]\textlambda$3427\,$\AA, [OII]\textlambda$3726$+$3729\,$\AA, [NeIII]\textlambda$3870\,$\AA,  H$\gamma$\textlambda$4341\,$\AA, [OIII]\textlambda$4386\,$\AA, H$\beta$\textlambda4861$\,$\AA, [OIII]\textlambda$4959\,$\AA, [OIII]\textlambda$5007\,$\AA, [NII]\textlambda$6548\,$\AA, H$\alpha$\textlambda$6563\,$\AA, [NII]\textlambda$6584\,$\AA.\footnote{For convenience, emission lines shall be referenced by species alone. [OIII] and [NII] shall be assumed to be [OIII]\textlambda$5007\,$\AA\ and [NII]\textlambda$6584\,$\AA{} unless otherwise specified.} 

Using the Peak Analysis utility (PAN)\footnote{https://www.ncnr.nist.gov/staff/dimeo/panweb/pan.html} written in Interactive Data Language we are able to obtain integrated relative spectral flux values for each emission line. After fitting a region of the spectrum, we run 500 bootstrapped Monte Carlo fits over the spectrum flux to constrain the error. For each emission line we fit a single narrow line component. We justify only using a single Gaussian given the high quality of fit and since we are only concerned with capturing the majority of flux rather than trying to fit the exact shape of the emission. However, there are two objects that showed clear visual evidence of blueshifted wings in the [OIII] emission that we examine in more detail in Section~\ref{sec:interesting}.

Following emission line fitting, we are able to generate a final redshift with its uncertainty for each object. A redshift is calculated for each emission feature in the objects with identifiable emission lines. Each object is assigned a redshift by calculating the weighted average of the emission line redshifts. The uncertainty is taken to be the PAN fitting uncertainties added in quadrature. The final redshifts are listed in Table~\ref{tab:fitdata}. We compare our measured spectroscopic redshifts against existing photometric redshifts from SDSS DR12 presented in \citet{Beck16drphot}. Only J035524 is not shown as \citet{Beck16drphot} was not able to generate a photometric redshift for this object. We note that our measured redshifts are consistent with the \citet{Beck16drphot} redshifts ($R^2=0.73$) and compare the two in Figure~\ref{fig:zcomp}.

In order to identify our objects as AGN, we utilise diagnostics developed using excitation line ratios. Our study will primarily make use of the \citet[hereafter TBT]{trou11optx} diagnostic with our sample. The TBT diagnostic compares optical rest frame $g-z$ colour ($^{0.0}(g-z)$) against the [NeIII]/[OII] emission ratio. The TBT diagnostic is consistent with the well-known \citet[hereafter BPT]{bald81bpt} diagnostic but also recovers a greater fraction of X-ray. Furthermore, the TBT emission lines remain in the optical regime out to $z<1.4$, whereas the BPT diagnostic's lines are limited to $z<0.5$. Given the limited spectral range of our spectra, we use TBT as it requires only one emission line ratio and has a greater redshift range for the optical wavelengths probed by our spectra.

Five objects with [OII] emission show no evidence of the [NeIII] emission feature. To obtain upper limits for these lines, we constrain a Gaussian to the predicted wavelength for the feature  given the redshift of the object. The full-width half maximum (FWHM) of the Gaussian is then set to the mean FWHM of the [NeIII] emissions in the sample. We then run the fitting procedure and take the upper limit on the error as our upper limit flux. Similarly, two objects with [OIII] emission showed no evidence of H$\beta$ emission. The upper limit flux for H$\beta$ was generated identically as described previously, however this results in a lower limit for the entire line ratio. All of the relevant emission line ratios we obtained for all of the objects are listed in Table~\ref{tab:fitdata}. 

We obtain the [NeIII]/[OII] emission line ratio or an upper limit on the ratio for 22 of our sample. The remaining objects have one or both of the lines outside of our spectral range or lying on a chip gap. Using optical rest frame $g-z$ colour ($^{0.0}(g-z)$) determined in Section~\ref{sec:SED}, we plot the objects on the TBT diagram to verify if they have AGN contribution. Figure~\ref{fig:TBT_contour} shows the SDSS galaxies verified as either AGN, star-forming, or composite using the BPT diagnostic with the demarcations outlined in \citet{kewl06agn} plotted in the TBT diagram. Over the SDSS contours we plot all of the objects with [NeIII] and [OII] measurements or [NeIII] upper limits. It is clear that all but two of these objects, J035524 and J204839, lie well within the AGN-Composite domain. 

Two of our objects, J081733 and J211845, have redshifts low enough for us to obtain [NII]/H$\alpha$ measurements along with [OIII]/H$\beta$. J081733, which has a $\log_{10}($[NII]/H$\alpha)$ of $-0.16\pm0.03$, and a $\log_{10}($[OIII]/H$\beta)$ of $+0.389\pm0.022$, lies in the area dominated by AGNs on the BPT diagram. The H$\beta$ for J211845 fell on a chip gap so we are unable to generate any value for this emission ratio. However, J211845 has a ratio of $\log_{10}($[NII]/H$\alpha)$ of $+0.13\pm0.01$, consistent with the ratios of AGNs on the BPT diagram.
\begin{figure}
	\centering
	\includegraphics[width=\columnwidth]{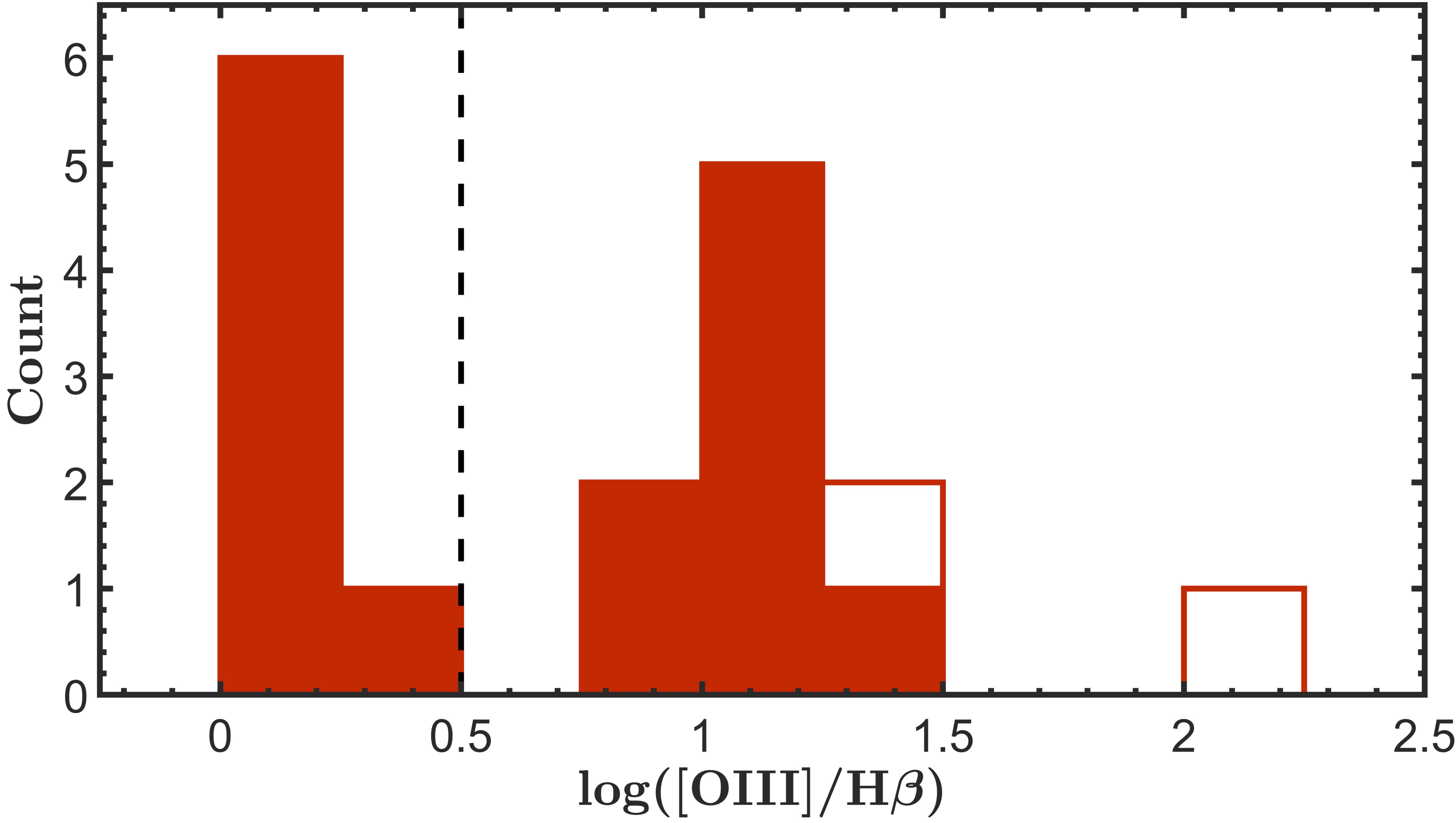}
	\caption{Distribution of log$_{10}$([OIII]/H$\beta$) for 17 objects with spectral coverage of this emission ratio. For two objects we were only able to obtain lower limits denoted with an unfilled box. The emission ratio AGN cutoff of 0.5 is also shown with more than half of the objects lying above this threshold.} 
    \label{fig:OIIIhist}
\end{figure}

\begin{figure}
	\centering
    \includegraphics[width=\columnwidth]{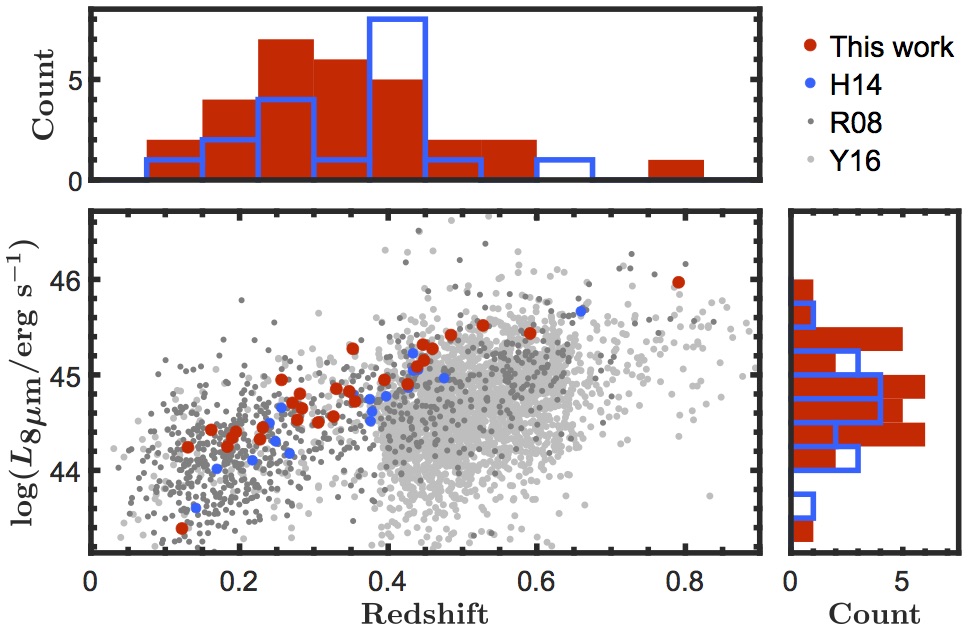}
	\caption{Intrinsic AGN 8$\,\mu$m luminosity against redshift along with histograms of each distribution on the axes. We include the SDSS Type II quasars from R08 in dark gray, the SDSSS Type II quasars from Y16 in light gray, and the `Group 1' objects from H14 with spectroscopic redshifts in blue. We plot the objects from this work in red.}
	\label{fig:LumZ}  
\end{figure}
\begin{figure}
	\centering
	\includegraphics[width=\columnwidth]{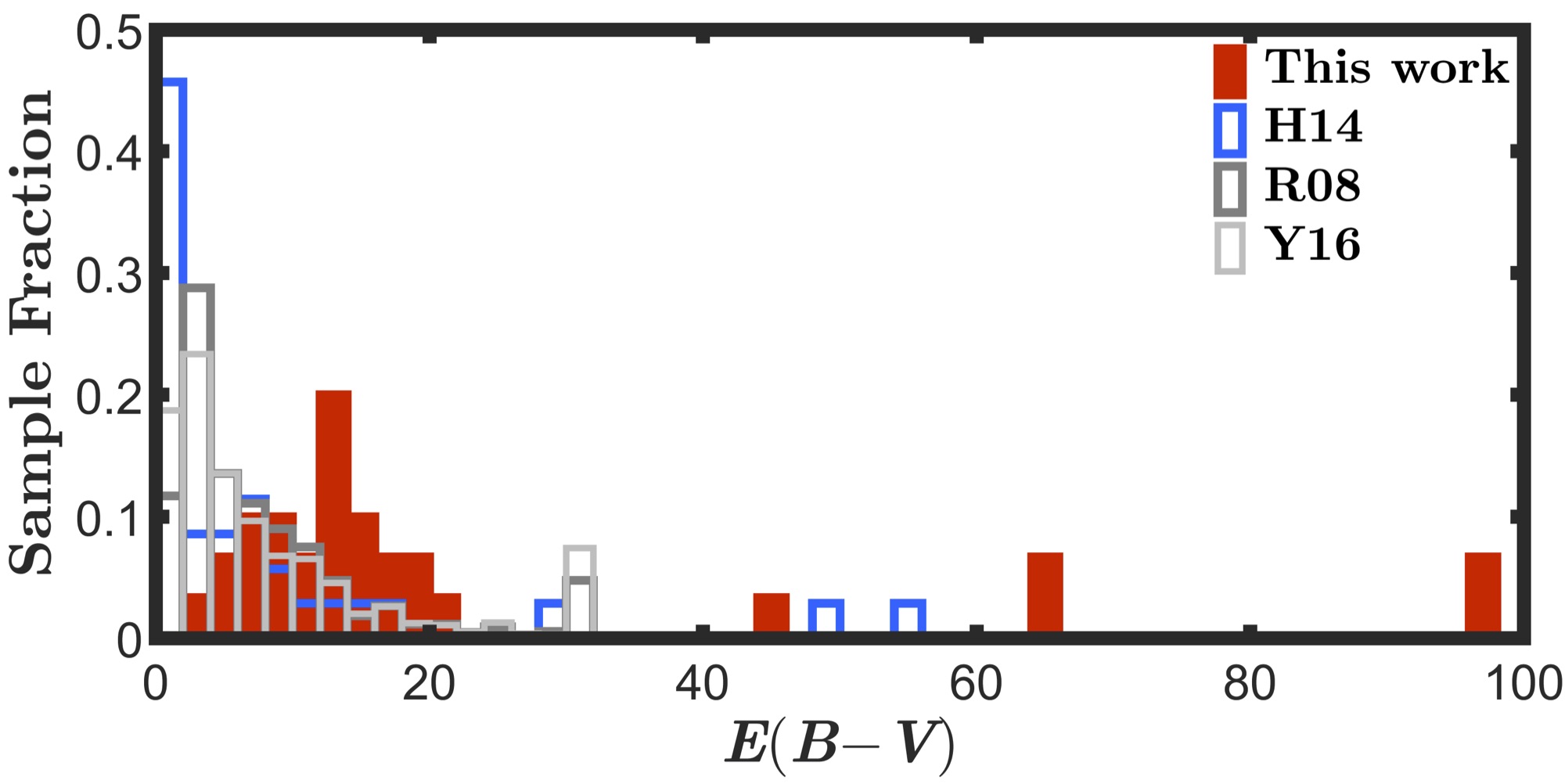}
    \caption{Fractional histogram of extinction coefficients of the quasar candidates from this work (red), H14 (blue), R08 (dark gray), and Y16 (light gray).}
	\label{fig:EBVdistrib} 
\end{figure}
We obtain the [OIII]/H$\beta$ ratio or a lower limit on the ratio for 17 objects in our sample. The full distribution of this ratio is shown in Figure~\ref{fig:OIIIhist}. Two objects in this distribution only have lower limit emission ratios depicted as empty boxes. Ten objects have $\log_{10}($[OIII]/H$\beta$) > 0.5, a ratio consistent with strong AGN emission, placing them in a regime primarily occupied by AGN. Any object satisfying this criterion is classified as an AGN for this study. This allows us to additionally identify J155048 as an AGN, which had its [NeIII] emission lie on a chip gap, whereas the remaining nine already lie in the AGN regime of the TBT diagram. While we do not have [NII]/H$\alpha$ ratios for these objects, their high [OIII]/H$\beta$ ratios are still strong indicators of AGN activity.

By combining our results from the TBT and BPT diagnostic, we can confidently place 21 (70\%) of our objects with identified redshift in the AGN regimes of one of these two excitation diagnostics. Of these, 11 (52\%) are identified solely using the TBT diagnostic, 1 (5\%) are identified solely using the BPT diagnostic, and the remaining 9 (43\%) are identified using both diagnostics. We stress that all but two of our objects (92\%) for which we have a line ratio are confidently placed in the AGN regime of one or more diagnostics. This supports the ability of our selection criteria to select AGN. No objects in our sample had spectral coverage that included all three emission line ratios. We are not able to place the remaining five objects (17\%) with identified redshift on either excitation diagram due to a lack of spectral coverage or chip gaps.
\begin{figure*}
	\centering
	\includegraphics[width=2.09\columnwidth]{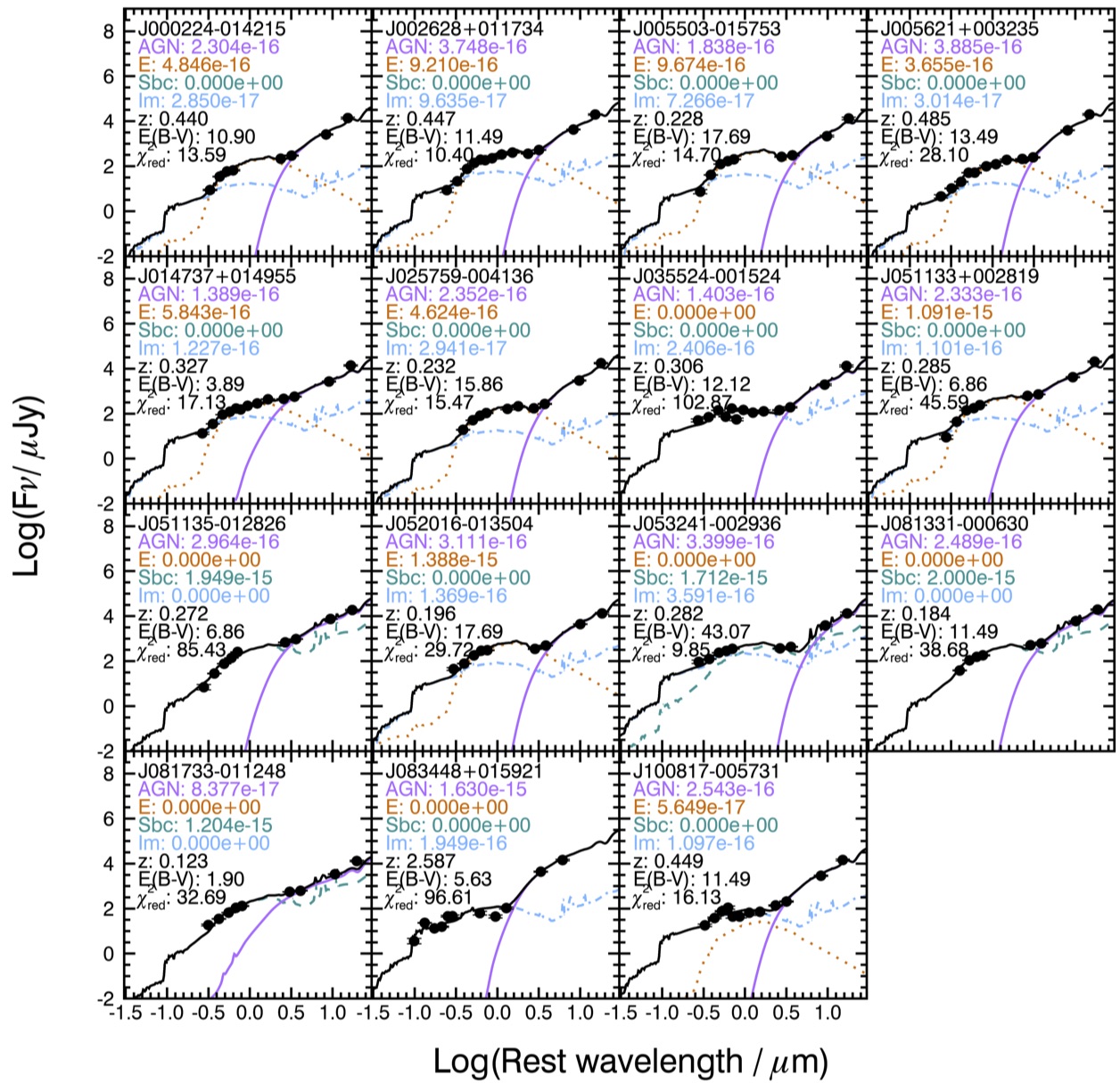}
    \caption{SED fits for the first 15 of our objects using templates from \citet{asse10agntemp}. Using a fixed redshift, the photometric data (black) are fit with composite SED generated from a quasar component (purple line), an elliptical component (orange dotted line), an irregular component (blue dot-dash line), a spiral component (green dashed line), and with extinction as another free parameter. The fitting results are detailed in Table~\ref{tab:fitdata}.}
	\label{fig:SEDFit}  
\end{figure*}
\begin{figure*}
	\centering
	\includegraphics[width=2.09\columnwidth]{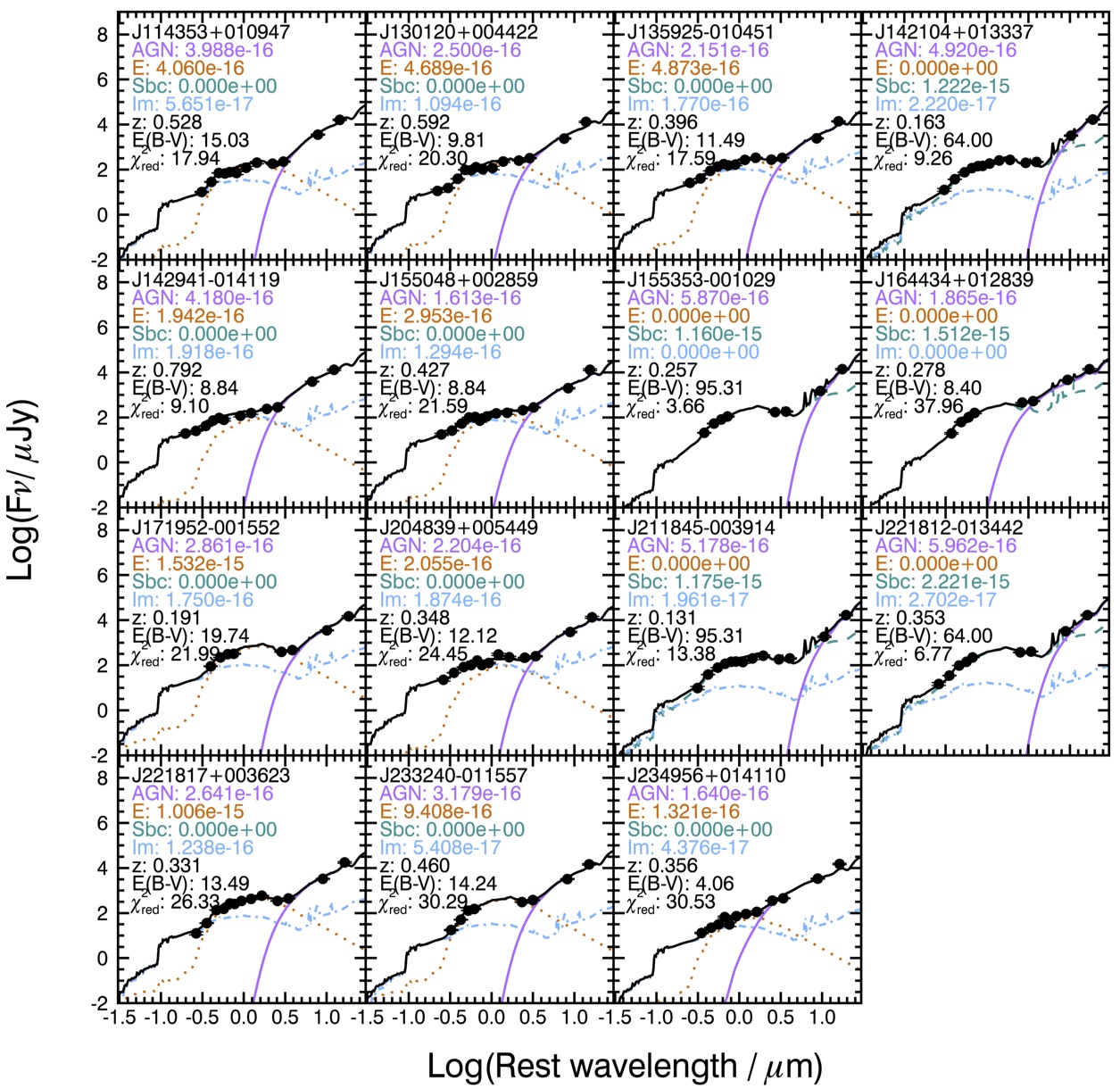}
	\caption{Continued from Figure \ref{fig:SEDFit}. SED fits for the last 15 of our objects using templates from \citet{asse10agntemp}. Using a fixed redshift, the photometric data (black) are fit with composite SED generated from a quasar component (purple line), an elliptical component (orange dotted line), an irregular component (blue dot-dash line), a spiral component (green dashed line), and with extinction as another free parameter. The fitting results are detailed in Table~\ref{tab:fitdata}.}
	\label{fig:SEDFit2}  
\end{figure*}
\begin{table*}
\begin{center}
  \caption{Sample Analysis Results}
  \label{tab:fitdata}
\setlength{\tabcolsep}{4.85pt}
  \begin{tabular}{ccccccccccr} 
  \hline \hline {SDSS Name}  & {$z_\textrm{spec}$} & {$^{0.0}(g-z)$} &{$[\text{NeIII}]/[\text{OII}]$}&{$[\text{OIII}]/\text{H}\beta$} & {$E(B-V)$} & {$f_\text{AGN,1$\,\mu$m}$} &\multicolumn{2}{c}{$f_\text{AGN,8$\,\mu$m}$} & {${L}_\text{8$\,\mu$m}/\text{erg s}^{-1}$} & {$\chi^2_\text{red} $}\\ & & & $^\text{a,\,b,\,c}$ & $^\text{a,\,b,\,d}$ & AGN & {int.}& {int.}& {obs.} & $^\text{a}$  \\
\hline 
J000224.02$-$014215.6&0.440&1.31$\pm$0.04&$-$0.64$\pm$0.41&1.058$\pm$0.003&10.90&0.45&0.99&0.98&45.36&13.59\\
J002628.56$+$011734.1&0.447&1.20$\pm$0.10&$-$0.14$\pm$0.07&1.457          &11.49&0.39&0.98&0.97&45.58&10.40\\
J005503.68$-$015753.3&0.228&1.27$\pm$0.02&$-$0.27$\pm$0.07&2.089          &17.69&0.24&0.97&0.93&45.09&14.70\\
J005621.72$+$003235.7&0.485&1.25$\pm$0.11&$+$0.03$\pm$0.05&1.008$\pm$0.002&13.49&0.63&0.99&0.99&45.62&28.10\\
J014737.42$+$014955.7&0.327&1.02$\pm$0.08&$-$0.38$\pm$0.15&---            &03.89&0.24&0.94&0.93&45.06&17.13\\
J025759.11$-$004136.9&0.232&1.30$\pm$0.04&$-$0.43$\pm$0.05&---            &15.86&0.46&0.99&0.98&45.20&15.47\\
J035524.81$-$001524.7&0.306&0.36$\pm$0.04&$-$0.76$\pm$0.09&---            &12.12&0.38&0.90&0.84&45.04&102.87\\
J051133.90$+$002819.0&0.285&1.21$\pm$0.02&$-$0.32$\pm$0.04&1.307$\pm$0.001&06.86&0.25&0.96&0.95&45.25&45.59\\
J051135.47$-$012826.7&0.272&1.33$\pm$0.03&$-$0.84         &0.017$\pm$0.021&06.86&0.37&0.73&0.68&45.34&85.43\\
J052016.18$-$013504.1&0.196&1.21$\pm$0.02&---             &0.178$\pm$0.013&17.69&0.26&0.97&0.93&45.29&29.72\\
J053241.28$-$002936.4&0.282&0.77$\pm$0.03&$-$0.25         &0.240$\pm$0.032&43.07&0.30&0.74&0.33&45.41&09.85\\
J081331.40$-$000630.6&0.184&1.33$\pm$0.03&---             &0.035$\pm$0.031&11.49&0.33&0.69&0.59&45.19&38.68\\
J081733.09$-$011248.7&0.123&1.35$\pm$0.04&---             &0.389$\pm$0.022&01.90&0.21&0.56&0.54&44.66&32.69\\
J083448.48$+$015921.1&2.587&0.36$\pm$0.26&---             &---            &05.63&0.90&0.99&0.99&47.18&96.61\\
J100817.25$-$005731.5&0.449&0.51$\pm$0.19&$-$0.26$\pm$0.26&0.825$\pm$0.006&11.49&0.65&0.97&0.96&45.41&16.13\\
J114353.77$+$010947.2&0.528&1.13$\pm$0.16&$-$0.40$\pm$0.03&---            &15.03&0.60&0.99&0.98&45.67&17.94\\
J130120.87$+$004422.6&0.592&0.99$\pm$0.20&$-$0.24$\pm$0.07&---            &09.81&0.41&0.97&0.95&45.51&20.30\\
J135925.18$-$010451.1&0.396&0.87$\pm$0.10&$-$0.40$\pm$0.10&0.914$\pm$0.005&11.49&0.33&0.95&0.92&45.30&17.59\\
J142104.04$+$013337.6&0.163&1.22$\pm$0.10&---             &0.091$\pm$0.022&64.00&0.60&0.88&0.35&45.46&09.26\\
J142941.46$-$014119.6&0.792&0.62$\pm$0.13&$-$0.56$\pm$0.11&---            &08.84&0.59&0.97&0.96&45.86&09.10\\
J155048.36$+$002859.7&0.427&0.82$\pm$0.11&---             &1.028$\pm$0.001&08.84&0.36&0.95&0.93&45.20&21.59\\
J155353.96$-$001029.3&0.257&1.33$\pm$0.03&$-$0.76         &---            &95.31&0.66&0.90&0.16&45.62&03.66\\
J164434.17$+$012839.5&0.278&1.33$\pm$0.05&$-$0.44         &0.190$\pm$0.056&08.40&0.33&0.69&0.61&45.14&37.96\\
J171952.75$-$001552.8&0.191&1.18$\pm$0.02&---             &---            &19.74&0.22&0.95&0.90&45.25&21.99\\
J204839.63$+$005449.3&0.348&0.64$\pm$0.10&$-$1.10         &---            &12.12&0.43&0.95&0.91&45.27&24.45\\
J211845.16$-$003914.6&0.131&1.23$\pm$0.09&---             &---            &95.31&0.62&0.89&0.14&45.45&13.38\\
J221812.68$-$013442.7&0.353&1.26$\pm$0.02&$-$0.45$\pm$0.17&---            &64.00&0.50&0.83&0.26&45.71&06.77\\
J221817.26$+$003623.6&0.331&1.16$\pm$0.08&$-$0.26$\pm$0.03&1.051$\pm$0.003&13.49&0.29&0.96&0.94&45.34&26.33\\
J233240.87$-$011557.9&0.460&1.31$\pm$0.03&$+$0.15$\pm$0.01&1.079$\pm$0.002&14.24&0.36&0.98&0.97&45.52&30.29\\
J234956.07$+$014110.2&0.356&0.90$\pm$0.06&$-$0.42$\pm$0.12&---            &04.06&0.59&0.98&0.98&45.15&30.53\\
\hline
  \multicolumn{10}{l}{int: intrinsic value with applied extinction correction, obs: observed value without an applied extinction correction.}\\
  \multicolumn{10}{l}{$^\text{a}$\,Values quoted are the log$_{10}$ of the measurement.}\\
  \multicolumn{10}{l}{$^\text{b}$\,Objects quoted without a value did not have spectral coverage of the lines, or had one of the lines lie on a chip gap.}\\
  \multicolumn{10}{l}{$^\text{c}$\,Objects quoted without an uncertainty are upper limits.}\\
  \multicolumn{10}{l}{$^\text{d}$\,Objects quoted without an uncertainty are lower limits}\\
  \end{tabular}
  \end{center}
\end{table*}
\section{Spectral Energy Distribution Models}
\label{sec:SED}

Given the existing SDSS and \textit{WISE} photometry for all of our objects, it is desirable to decompose the SEDs of our objects with measured redshift in order to understand the contribution of the AGN to the total SED, and the extinction of the AGN emission. Throughout this section, objects are modeled following the procedure outlined in Carroll et al. (in prep).

We fit the sample using the four empirically-determined \citet{asse10agntemp} galaxy\footnote{Spiral (Sbc), elliptical (Elip), and irregular (Im) templates used.} and AGN templates. Objects are fit using optical dereddened model magnitudes from SDSS Data Release 12 \citep{SDSSDR11+12}, mid-IR photometry from allWISE, and available near-infrared petrosian magnitudes from the United Kingdom InfraRed Telescope Infrared Deep Sky Survey Data Release 10 \citep{LawrenceUKIDSS}. Using the spectroscopically-determined redshifts found in Section~\ref{sec:reduction}, we shift the templates to the observed frame of our objects. Following \citet{asse10agntemp}, we apply an extinction model to the AGN template in order to simulate the effects of AGN obscuration. The extinction consists of a Small Magenellenic Cloud like extinction curve at short wavelengths \citep[\textlambda$ < 3300$\,\AA]{Gordon98SMC} and a Galactic extinction curve otherwise \citep{Cardelli89Dust}. Throughout this work we parametrise the extinction using $E(B-V)$ and assume $R_V=3.1$ for both extinction curves. We create a grid of 100 exponentially spaced extinction coefficient ($E(B-V)_\textrm{AGN}$) values from 1 to 95 and apply them to the AGN template. We perform a linear-least-squares fit to the available photometry using the templates, and use the fit with the lowest reduced $\chi^2$ as the solution. We then extract the rest-frame magnitudes from our best template fit, convolving the coadded template with the instrument transmission curves. The error in rest frame magnitudes is approximated by finding the percent error in the observed frame photometry measurement nearest to the redshifted rest frame band wavelength. The final error in the $^{0.0}(g-z)$ colour is taken as the $g$ and $z$ errors added in quadrature. The relevant results from our SED fitting are enumerated in Table ~\ref{tab:fitdata}.

In order to assess the AGN contribution of our sample we calculate the intrinsic luminosity of the AGN. We aim to find the rest-frame 8$\,\mu$m luminosity of our sample in order to be consistent with H14. 8$\,\mu$m corresponds to an observed frame wavelength of 12$\,\mu$m (assuming $z\sim 0.5$), where the sensitive $W3$ band lies, whereas longer wavelengths would correspond to less \textit{WISE} sensitivity. Given that the \citet{asse10agntemp} templates include PAH emission, we expect minimal error arising from this kind of galaxy contamination. In addition, the emission in rest frame 8$\,\mu$m is dominated by the AGN power law dominates and has less significant galaxy contribution. To highlight this distinction, we calculate the AGN fraction ($f_\text{AGN}$: the AGN contribution over the total SED) at two wavelengths, 1 and 8$\,\mu$m. We also calculate the AGN fraction in two cases, an intrinsic value with applied extinction correction generated from the fitted extinction coefficient values, and an observed value without the correction. The $f_\text{AGN}$ at 1$\,\mu$m was heavily affected by extinction where the average intrinsic $f_\text{AGN}$ was 0.38 while the average uncorrected $f_\text{AGN}$ was 0.00. The $f_\text{AGN}$ at 8$\,\mu$m was almost unaffected by extinction where the average intrinsic $f_\text{AGN}$ was 0.91 while the average uncorrected $f_\text{AGN}$ was 0.80. 

In conjunction with luminosity distances obtained from spectroscopic redshifts, we therefore use the rest-frame 8$\,\mu$m flux obtained from the SED fitting in order to generate rest-frame 8$\,\mu$m luminosities ($L_{8\,\mu\text{m}}$) for all of our objects. We additionally run the entire fitting procedure on the objects in Y16 and R08 and make use of the values from `Group 1' in H14. In Figure~\ref{fig:LumZ} we plot the $\log(\text{L}_{8\,\mu\text{m}}/\text{erg}\,\text{s}^{-1})$ against the spectroscopically determined redshift for all four samples along with histograms for each axis. Our sample, similar to the `Group 1' in H14, are generally more luminous than the AGNs found in Y16 and R08 at a given redshift. This is expected as our mid-IR $W4$ selection criterion was chosen specifically to select bright AGNs in a certain luminosity range and is identical to a H14 `Group 1' criterion. Furthermore, this work and 'Group 1' from H14 occupy a similar range in redshift space, likely a by-product of identical $g$ band magnitude cut on the samples. 

Objects in the sample are found to exhibit large extinction coefficients, with an average $E(B-V)_\textrm{AGN}$ of 17.8 corresponding to an average $A(V)_\textrm{AGN}$\ of 51.4. Figure~\ref{fig:EBVdistrib} shows a fractional histogram of extinction values for our sample, `Group 1' from H14, Y16, and R08. Our objects exhibit much higher average extinction coefficients than the other surveys, indicating that we are indeed probing a more heavily obscured population. Specifically, we highlight the increased extinction levels relative to H14, whose `Group 1' objects formed the basis for this study. Given that the samples have near identical selection criterion, we attribute the increased level of obscuration to the requirement to lie below the Mateos wedge the  likely comes from this work probing a region of \textit{WISE} colour space that is populated with these heavily obscured quasars. 

However, we note that the use of the extinction parameter, modeled as a simple screen, does not represent the real physics of the obscuring material. Studies, such as \citet{Nenkova08Tor}, have demonstrated that exact extinction is controlled by many parameters, rather than a simple correction factor. While the $E(B-V)$ coefficient is a useful tool in controlling the shape of the mid-IR shape of the AGN component, we reiterate that it cannot be interpreted as an immediate translation to the exact geometry of the intrinsic obscuration.

Figures~\ref{fig:SEDFit} and~\ref{fig:SEDFit2} show all of the SED fits for our objects with the \citet{asse10agntemp} galaxy and AGN templates colour coded with relevant fitting output. Our fitting results reinforce the need for infrared colour selection of these objects where the AGN contribution is a larger fraction of the total coupled with high levels of obscuration. Our SED modeling emphasises that we are indeed probing a population of extremely luminous, heavily obscured quasars. 

\section{Composite Spectrum}
\label{sec:composite}
In order to compare our sample to similar type II quasar samples with composite spectra such as Z03 and L13, we generated a composite spectrum using our objects with more than one identifiable emission line. We generated a composite spectrum for `Group 1' objects from H14 to provide an additional comparison.
\begin{figure*}
	\centering
	\includegraphics[width=2.09\columnwidth]{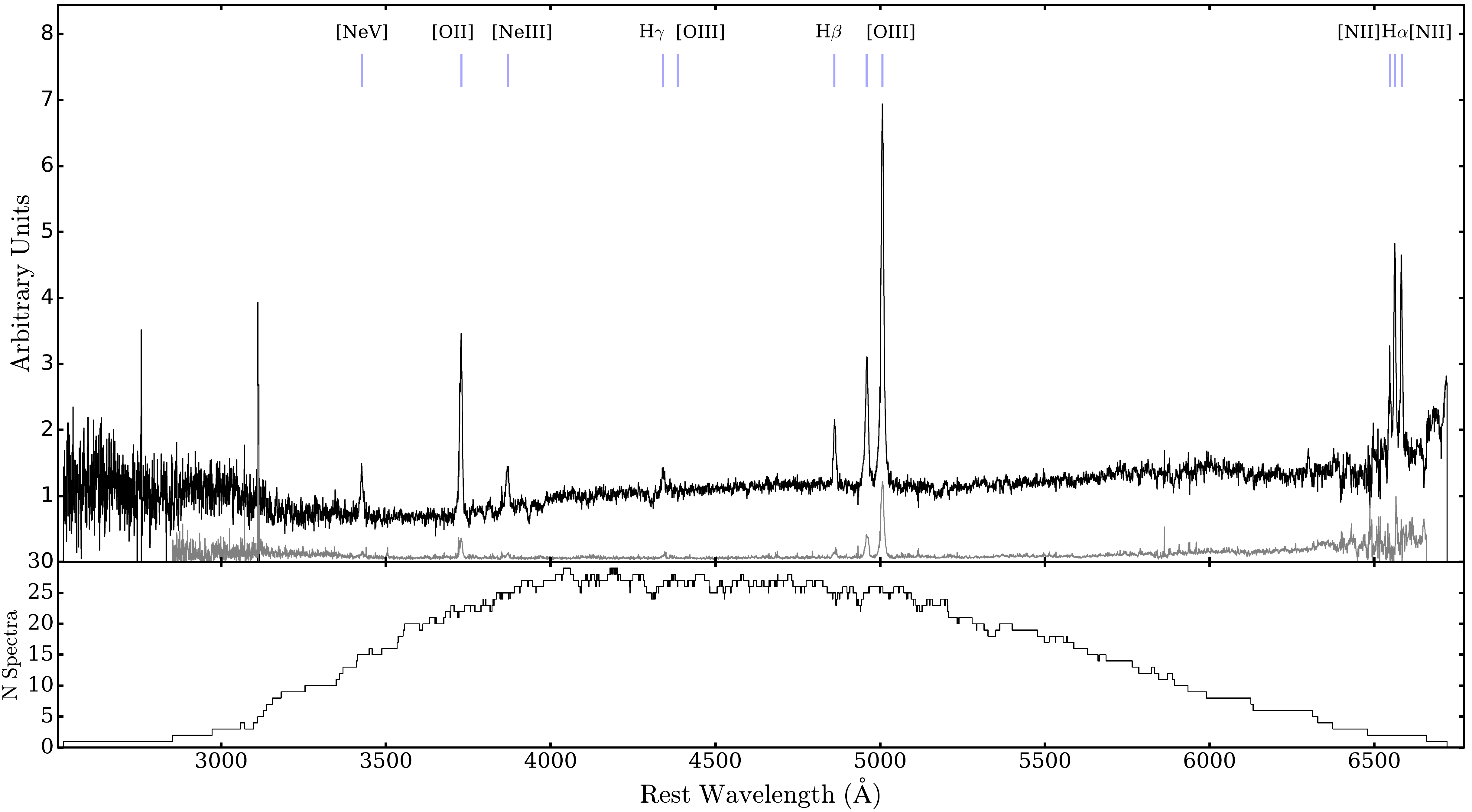}
	\caption{The composite spectrum for the objects in this sample with two or more identified emission lines generated by normalizing all of the objects to an area of undisturbed continuum. Emission lines are marked by blue lines and are labeled at the top of the page. The uncertainty spectrum is plotted in gray. The number of contributing points per wavelength is also shown. 
	\label{fig:compspec}
   	}
\end{figure*}
\begin{figure*}
	\centering
	\includegraphics[width=2.09\columnwidth]{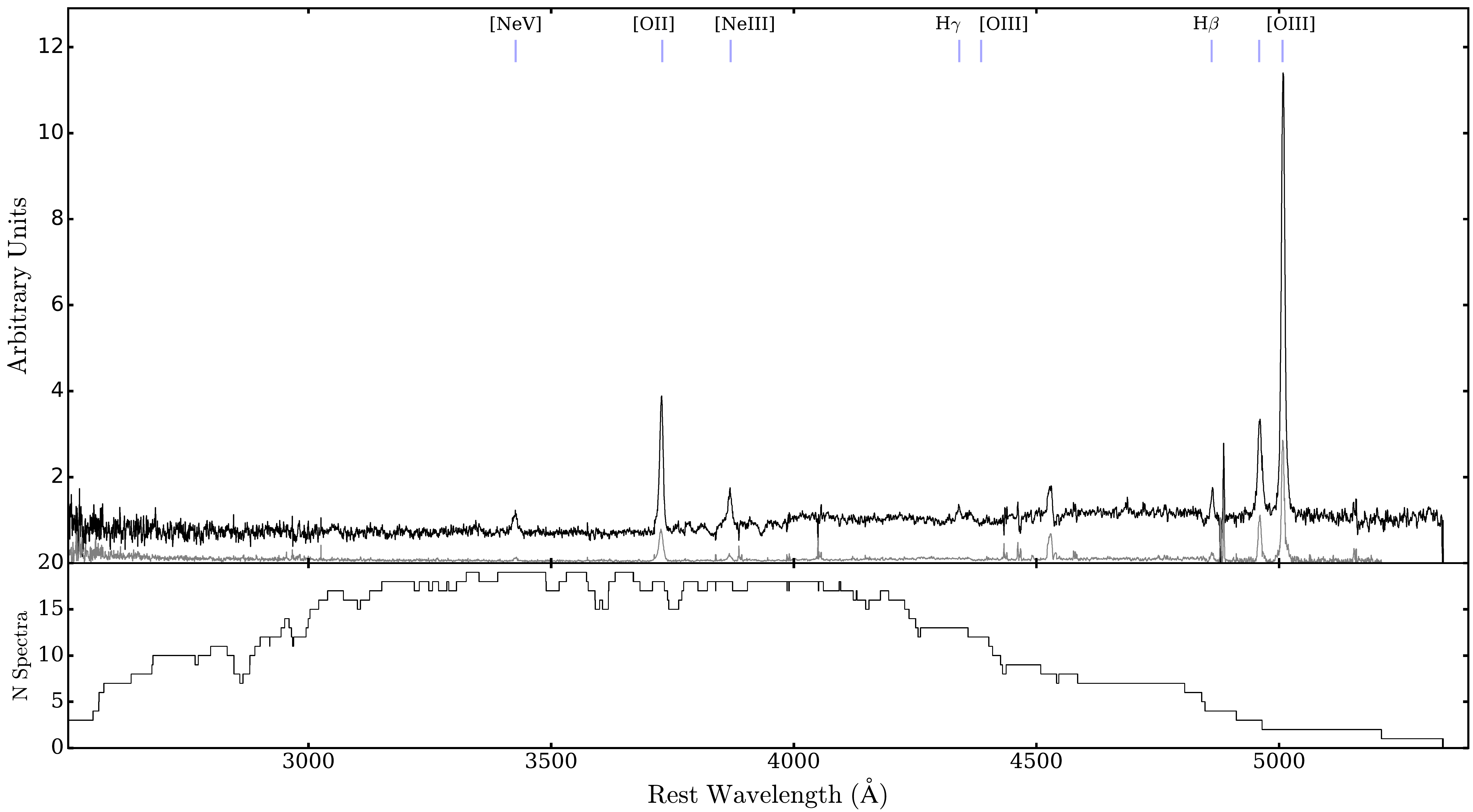}
	\caption{The composite spectrum for the objects in H14 generated by normalizing all of the objects to an area of undisturbed continuum. Emission lines are marked by blue lines and are labeled at the top of the page. The uncertainty spectrum is plotted in gray. The number of contributing points per wavelength is also shown. 
	\label{fig:compspeckev}
   	}
\end{figure*}
\begin{figure*}
	\centering
	\includegraphics[width=2.09\columnwidth]{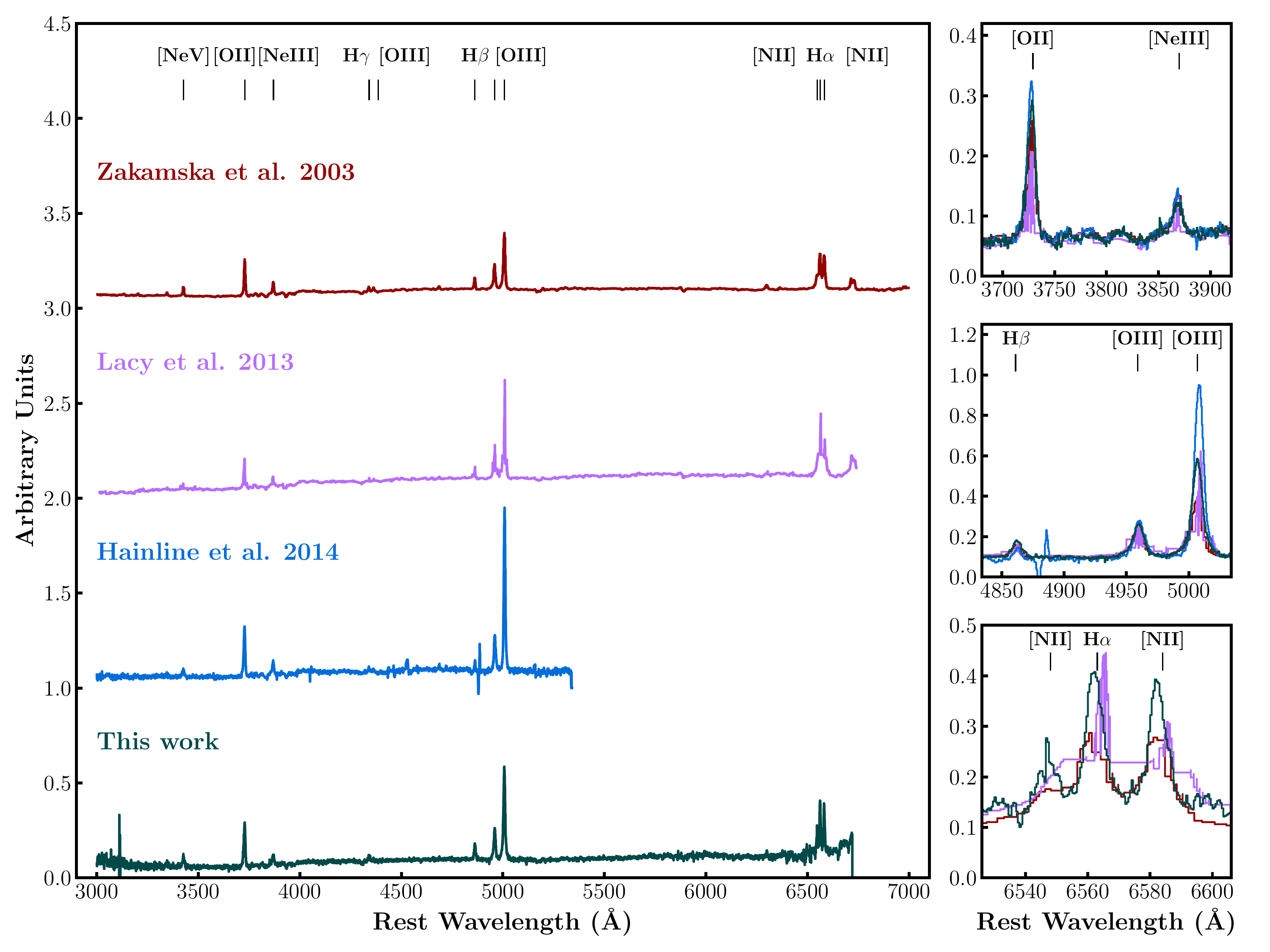}
	\caption{Composite spectra for this work, H14, L13, and Z03 presented for comparison. Each spectrum is normalised to the same area of the continuum from 4150 to 4250\AA. Relevant emission lines are presented in side-panels and their ratios are presented in Table~\ref{tab:compdata}.}
    \label{fig:allcomp} 
\end{figure*}

\begin{table}
\setlength{\tabcolsep}{2.45pt}
\begin{center}
  \caption{Comparison of Composite Spectra}
  \label{tab:compdata}
  \begin{tabular}{lccc} 
  \hline \hline {Comp. Spectra} & \multicolumn{3}{c}{Log$_{10}$ of Emission Line Ratio} \\ 
  & {[NeIII]/[OII]} &{[OIII]/H$\beta$} & {[NII]/H$\alpha$} \\
\hline This work	&	$-$0.46	$\pm$	0.07	&	0.756	$\pm$	0.005	&	$-$0.10	$\pm$	0.05	\\
Hainline et al. 2014$^\text{a}$	&	$-$0.41	$\pm$	0.06	&	1.456	$\pm$	0.002	&	---			\\
Lacy et al. 2013$^\text{b}$	&	$-$0.48			&	1.03			&	$-$0.22			\\
Zakamska et al. 2003	&	$-$0.46	$\pm$	0.01	&	0.740	$\pm$	0.006	&	$+$0.03	$\pm$	0.02	\\ \hline
  \multicolumn{4}{l}{$^\text{a}$\,The Hainline et al. 2014 composite does not cover NII/H$\alpha$.}\\
  \multicolumn{4}{l}{$^\text{b}$\,Lacy et al. 2013 does not quote flux uncertainties.}\\
  \end{tabular}
  \end{center}
\end{table}

\begin{table*}
\begin{center}
  \caption{Analysis for Objects without Identified Emission}
  \label{tab:nospec}
\setlength{\tabcolsep}{11pt}
  \begin{tabular}{clcccccr} 
  \hline \hline {SDSS Name}  & {$z_\textrm{phot}$} & {$E(B-V)$} & {$f_\text{AGN,1$\,\mu$m}$} &\multicolumn{2}{c}{$f_\text{AGN,8$\,\mu$m}$} & {${L}_\text{8$\,\mu$m}/\text{erg s}^{-1}$} & {$\chi^2_\text{red} $}\\ &  & AGN & {int.}& {int.}& {obs.} & $^\text{a}$  \\
\hline 
J000956.62$-$002713.5 & 0.40            & 10.34 & 0.40 & 0.97 & 0.95 & 44.16 & 17.86\\
J045401.15$-$003822.2 & 0.32            & 15.86 & 0.39 & 0.75 & 0.61 & 44.17 & 20.77\\
J053056.03$-$010012.5 & 0.85$^\text{b}$ & 09.81 & 0.59 & 0.99 & 0.99 & 45.88 & 326.58\\
J061340.30$-$005119.8 & 0.17            & 07.98 & 0.17 & 0.48 & 0.40 & 43.34 & 86.98\\
J100848.15$+$011801.4 & 0.34            & 24.61 & 0.84 & 0.99 & 0.99 & 44.52 & 05.70\\
J113954.32$-$010500.9 & 0.19            & 95.31 & 0.70 & 0.92 & 0.19 & 44.16 & 03.68\\
J131015.79$-$010321.6 & 0.31            & 20.85 & 0.69 & 0.99 & 0.98 & 44.45 & 05.26\\
J150858.17$+$004314.0 & 0.45            & 13.49 & 0.52 & 0.84 & 0.75 & 44.70 & 25.15\\
J154826.03$+$004615.3 & 0.46            & 03.72 & 0.23 & 0.57 & 0.54 & 44.16 & 28.97\\
J154909.79$+$011940.5 & 0.35            & 01.00 & 0.31 & 0.86 & 0.86 & 43.69 & 57.12\\
J162201.42$+$002931.8 & 0.29            & 10.90 & 0.35 & 0.72 & 0.62 & 44.06 & 137.16\\
J180408.11$+$010004.0 & 2.93$^\text{b}$ & 05.91 & 0.63 & 0.92 & 0.90 & 47.35 & 65.23\\
J195611.27$-$000718.0 & 0.38            & 17.69 & 0.40 & 0.75 & 0.60 & 44.45 & 36.80\\
J212649.41$-$000257.7 & 0.43            & 51.02 & 0.83 & 1.00 & 0.97 & 44.87 & 01.89\\
J223059.01$-$000057.5 & 0.18            & 10.34 & 0.16 & 0.47 & 0.37 & 43.50 & 24.61\\
\hline
  \multicolumn{8}{l}{int: intrinsic value with applied extinction correction, obs: observed value without an applied extinction correction.}\\
  \multicolumn{8}{l}{$^\text{a}$\,Values quoted are the log$_{10}$ of the measurement.}\\
  \multicolumn{8}{l}{$^\text{b}$\,Photometric redshift taken from the XDQSO catalogue.}\\

  \end{tabular}
  \end{center}
\end{table*}

We linearly interpolate the spectra on a grid of half-angstrom intervals, in order to avoid undersampling, using the \citet{Carnall17SpectRes} \texttt{SpectRes} python tool. Our entire sample does not overlap in an area of continuum without emission lines, chip gaps, or sky features. We therefore pursue normalizing the spectra to one another in the following fashion: we split the objects into three groups that do have overlapping undisturbed continuum ($z > 0.5$, $0.5 > z > 0.3$, $0.3 > z$ for this sample; $z > 0.3$, $0.3 > z > 0.175$, $0.175 > z$ for the H14 sample). We then normalise each spectrum in each group with respect to one another using an area of undisturbed continuum ($3125-3275\,$\AA, $3500-3600\,$\AA, $4100-4200\,$\AA\ for this work; $3150-3250\,$\AA, $3500-3600\,$\AA, $4000-4100\,$\AA\ for H14). To create a composite spectra for each group, we then use the sigma-clipping procedure in NumPy to iteratively eliminate outliers in each redshift bin for each spectrum. The value of each composite spectrum in a particular wavelength bin is taken to be the mean of the group data points remaining after the sigma-clipping while the uncertainty is taken to be the standard deviation of the remaining group data points divided by the square root of the number of remaining points. The three composite spectra are normalised with respect to each other using an area of continuum ($3500-3600\,$\AA). The normalization factors are applied to the original set of corresponding spectra, ensuring that the total set of spectra are properly normalised. Finally, the total composite spectrum is generated in the fashion outlined earlier but now with all of the spectra properly normalised. The smoothed composite spectrum for this sample is plotted in Figure~\ref{fig:compspec} along with the uncertainty and the number of contributing data points while the same is shown for the H14 sample in Figure~\ref{fig:compspeckev}.

The comparison of emission line ratios for our composite spectrum to the Z03, L13, and H14 composites is presented in Table~\ref{tab:compdata} along with each composite presented alongside each other in Figure \ref{fig:allcomp}. Specifically we make use of the variance-weighted composite spectrum from Z03, in order to preserve the relative fluxes of emission lines. Between all four studies, the value of the [NeIII]/[OII] ratio is consistent. Our value of the [OIII]/H$\beta$ ratio is consistent with the Z03 value. Our value is likely inconsistent with H14 as this line ratio is based only on two contributing spectra and is therefore an unreliable summary of all objects in the H14 sample. Similarly, for the [NII]/H$\alpha$ ratio, our value is based only on two contributing spectra and is therefore an unreliable summary of our sample. The similarity between the emission line ratios between our composite and the others again reinforces that our sample is comprised of type II AGN. 

\section{Objects without Identified Emission}
\label{sec:noemiss}
For eleven (34\%) of our objects we were not able to measure a redshift, due to either the lack of spectral features, or an inability to identify a single isolated emission line. In order to characterise this significant portion of our sample, we model the SEDs of these objects. The objects are fit in a process identical to Section~\ref{sec:SED} where a fixed redshift is assumed taken from the \citet{Beck16drphot} photometric redshifts for the SDSS DR12 catalogue supplemented with the \citet{dipo15qsoz} SDSS XDQSO photometric redshifts. We present the relevant results from our SED fitting in Table \ref{tab:nospec}.

In general, the results of the model fitting are consistent across objects with and without identified emission. Objects without emission had a lower average $f_{\textrm{AGN},8\mu m}$ AGN of 0.81, compared to 0.91 for the rest of the sample, suggesting this subset has fewer powerful quasars than objects with identified emission. Secondly, we find a higher average extinction coefficient of $E(B-V)=18.9$. 

In sum, we hypothesise that this subset may be partly comprised of powerful AGN whose emission features are extincted by the intervening material to the extent that they would be undetectable through our observations. Given the assumed photometric redshifts of the subset, the theorised extinction would attenuate both the [NeIII]/[OII] and [OIII]/H$\beta$ ratios. While we cannot confirm any of these objects as AGN through their optical spectroscopy, they present the potential to be AGN with obscuration severe enough to attenuate potential X-ray and optical emission and would therefore be recoverable only through mid-IR colour selection. 
 
\begin{figure}
	\centering
	\includegraphics[width=\columnwidth]{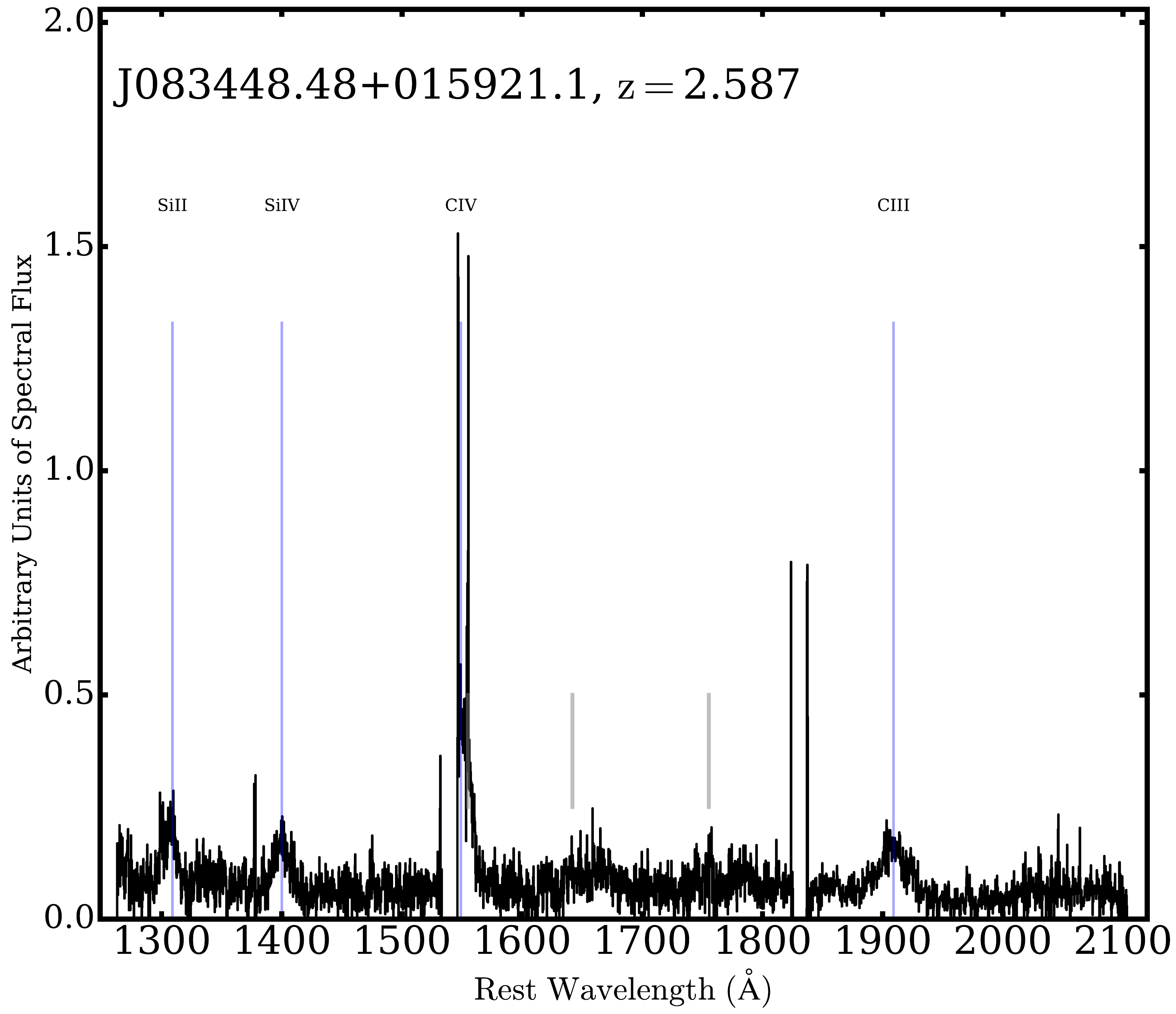}
	\caption{SALT long-slit observed spectroscopy of J083448, our only high redshift source. The object is plotted with total flux normalised to unity. Telluric lines are denoted with with vertical gray bars and areas of high atmospheric disturbance with horizontal gray bars.}
	\label{fig:HiZQSO}
\end{figure}
\begin{figure}
	\centering
	\includegraphics[width=\columnwidth]{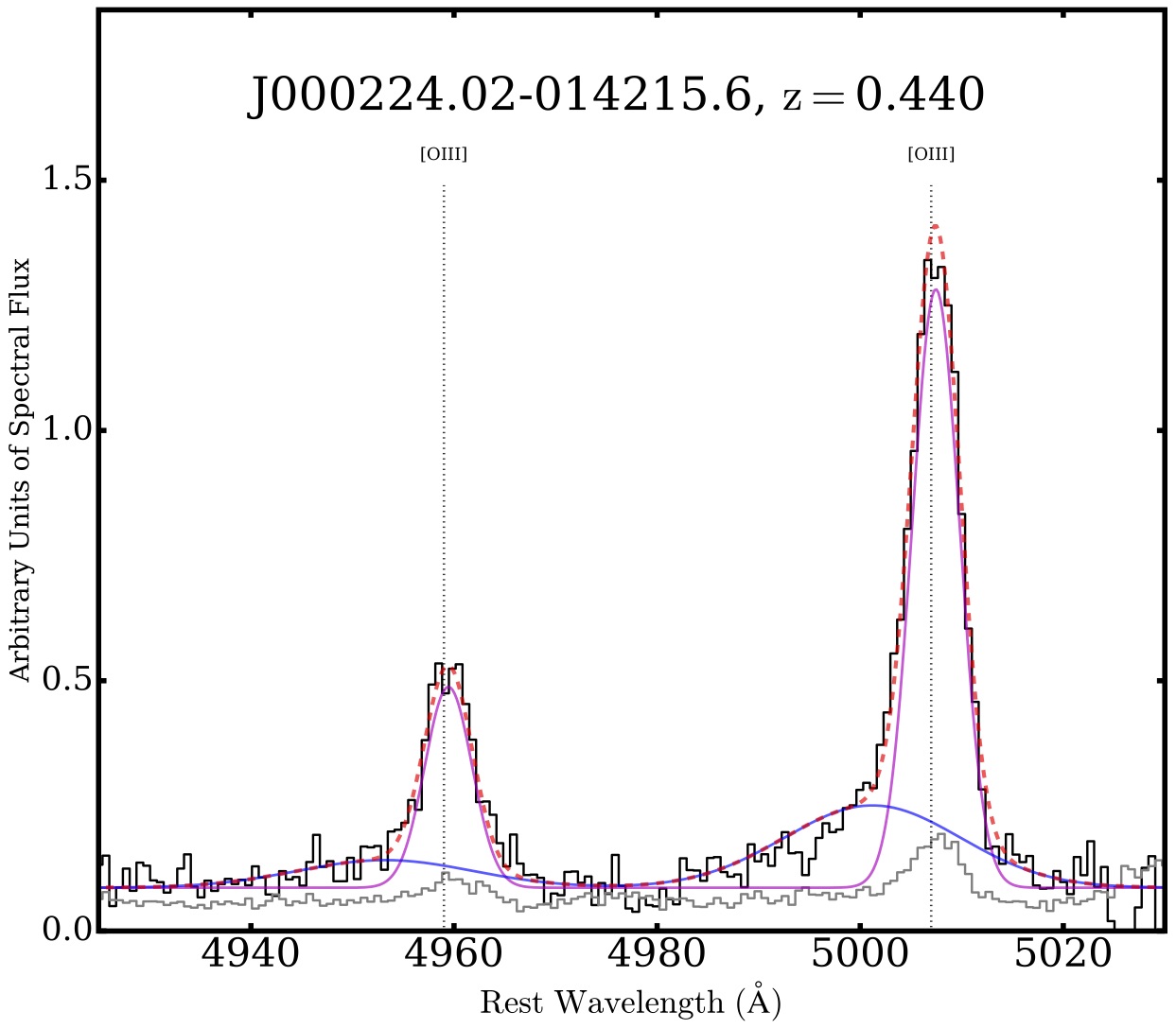}
	\caption{Rest-frame long-slit SALT spectroscopy of the [OIII]\textlambda$5007\,$\AA\ and [OIII]\textlambda$4959\,$\AA\ emission of J000224. The spectrum exhibits blueshifted wings with an offset of $291\,\text{km}\,\text{s}^{-1}$ and a broad component FWHM of 1227\,km\,s$^{-1}$. The narrow components are plotted in pink, the broad components in blue, and the overall fitting result as a dashed red line. The uncertainty spectrum is plotted below in gray.}
     \label{fig:J000224}
\end{figure}
\begin{figure}
	\centering
	\includegraphics[width=\columnwidth]{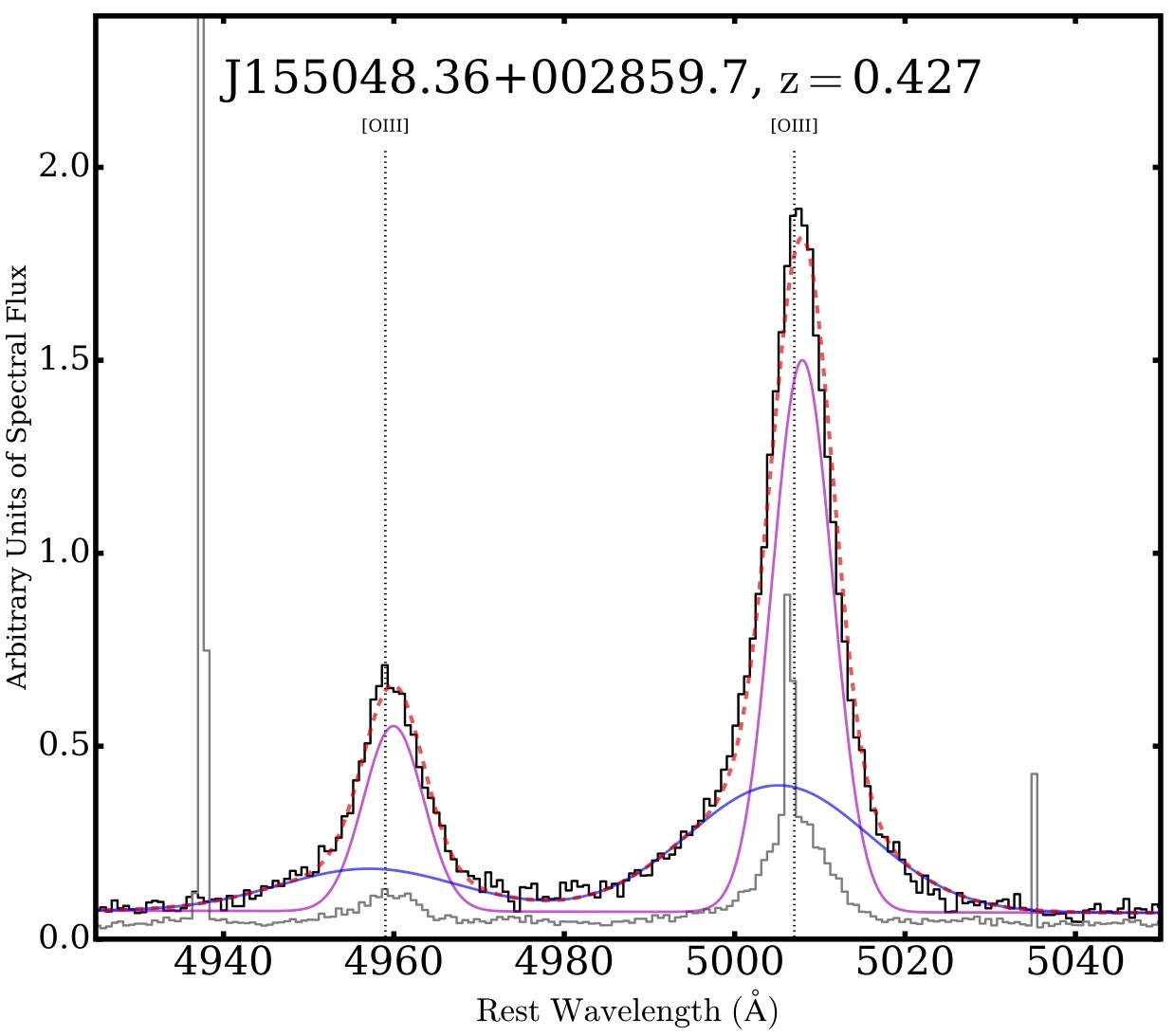}
	\caption{Rest-frame long-slit SALT spectroscopy of the [OIII]\textlambda$5007\,$\AA\ and [OIII]\textlambda$4959\,$\AA\ emission of J155048. The spectrum exhibits blueshifted wings with an offset of $168\,\text{km}\,\text{s}^{-1}$ and a broad component FWHM of 1506\,km\,s$^{-1}$. The narrow components are plotted in pink, the broad components in blue, and the overall fitting result as a dashed red line. The uncertainty spectrum is plotted below in gray.}
    \label{fig:J155048} 
\end{figure}

\section{Individual Targets}
\label{sec:interesting}
From our survey we identify several sources that merit further study and possibly deserve follow-up observations to determine additional spectral properties. 

Figure~\ref{fig:HiZQSO} shows the spectrum of J0883448, our only high-redshift source $(z=2.59)$. We conclude that J0883448 is a high redshift source as we are able to identify the emission features to be SiII\textlambda$1309\,$\AA, SiIV\textlambda$1400\,$\AA, CIV\textlambda$1549\,$\AA, and CIII\textlambda$1909\,$\AA. However, we note that the SED fit for this object, presented in Figure \ref{fig:SEDFit}, is likely inappropriate. The existing rest-frame UV broad lines in the spectrum do not originate primarily from starlight, as suggested by the SED. It may be of interest to examine the SED of this object in more detail in the future.

H14 also find two high-redshift quasars in the same redshift range. The contribution of these high redshift sources is small, with only one out of 30 objects with redshift in this work, and only two out of 35 objects with redshift in H14. 

We also present two objects, J000224 and J155048, with the clearest visual evidence of blueshifted wings in the [OIII]\textlambda5007\,\AA\ and [OIII]\textlambda4959\,\AA\ emission lines. Using PAN, we fit four Gaussians to the two lines, each with a broad and narrow component. We constrained the FHWM of the narrow components to be identical; likewise for the broad components. Similarly, the separation between the narrow components and broad components was set to be 48$\,$\AA, the difference between 5007$\,$\AA\ and 4959$\,$\AA. Finally, the area of the Gaussian for the [OIII]\textlambda4959\,\AA\ emission was set to be 1/2.98 the area of the [OIII]\textlambda5007\,\AA\ Gaussian, the fixed quantum mechanical ratio between these two lines described in \citet{Galavis97}. The results of our fitting for J000224 and J155048 are shown in Figures~\ref{fig:J000224} and~\ref{fig:J155048} respectively with the narrow and broad components colour-coded. 

In J000224 the broad component is offset by a velocity of 291$\,\text{km}\,\text{s}^{-1}$ with a FWHM of 1227$\,\text{km}\,\text{s}^{-1}$. In J155048 the broad component is offset by a velocity of 168$\,\text{km}\,\text{s}^{-1}$ with a FWHM of 1506$\,\text{km}\,\text{s}^{-1}$. These values are roughly consistent with similar spectroscopic studies of [OIII] emission \citep[e.g.][]{Vega01Kinem, Mullaney13Kinem, Collet16feedback, Harrison16KASHz, Sun17Kinem}. We note that most of these works cite the $w_{80}$ measurement for their objects, the velocity difference between the 10th and 90th percentiles of the emission line, roughly the FWHM of a single Gaussian fit. Our broad component FWHM therefore overestimate these measurements but likely still satisfy the $w_{80}>600\,\text{km}\,\text{s}^{-1}$ criterion outlined in \citet{Collet16feedback} and \citet{Liu13feedback} indicative of either ionised outflows or highly turbulent material. These objects are potentially of interest for further study with highly resolved imaging to understand their host properties and the location of the turbulent or outflowing material. 

\section{Summary and Conclusions}
\label{sec:conclusion}
From our sample of 46 target candidate obscured AGN, 30 exhibit emission lines that allowed us to determine redshifts, one had a stellar spectrum, and the remaining 15 had no identifiable features or one unidentifiable emission line. Of the 30 objects with redshift, all required a strong AGN component in the SED fits. Twenty-one objects (70\%) of objects with one or more identified emission features were placed confidently in the AGN regime on either the TBT or BPT diagram. These optically confirmed quasars lie outside infrared cuts based on X-ray selection and require a large optical-to-infrared extinction component in our models, suggesting that there exists a population of highly obscured quasars that are missing from mid-IR selections based on X-ray criterion and therefore perhaps missed in X-ray surveys. In a 3216$\,$deg$^2$ area of the sky, the \citet{mate12xmmwise} selects 2009 sources when subjected to our $g$ band and $W4$ flux cuts, compared to the 660 objects selected by our colour criteria.  This potentially represents nearly a quarter of missed AGN in X-ray selected studies. In fact, these objects may only be recoverable using mid-IR colour selection.

It is worth noting that our selection criteria were quite restrictive, especially in the \textit{W4} band, in order to preferentially select objects with bright mid-IR to find luminous obscured quasars. Given that most of our redshift identified sample were found to be optically confirmed AGN, we theorise that there may be other AGN sources with lower luminosities relative to their host galaxy that do not satisfy our photometric cut. The SED modeling of our redshift unidentified objects additionally hints at this even fainter population of obscured quasars, which would also be missing from X-ray surveys. 

There are several projects that follow naturally from our results. Our survey targets were motivated in part by non-detections in the X-ray band, and therefore we would target our objects with deeper observations with \textit{NuSTAR} that may be required to recover any X-ray emissions that penetrate the obscuring material. This may also be a project that is better suited for the upcoming European Space Agency space observatory, Advanced Telescope for High-ENergy Astrophysics \citep[ATHENA]{barret13athena}, or the proposed NASA space observatory, Lynx \citep{Weisskopf-Surveyor}.

Furthermore, it would be of interest to use imaging to resolve galactic features in order to locate the source of the obscuring dust in the galaxy and its relation to the central quasar engine. This project would be well suited for the low redshift objects in our sample. In addition, several objects showed evidence of blueshifted wings and outflowing or turbulent material, motivating a follow up with spatially resolved spectroscopy to determine kinematics of the system. 

\section*{Acknowledgements}

We would like to thank the anonymous referee for their constructive comments which improved the final paper. This work uses data taken from the Southern African Large Telescope, and it is supported in part by the The James O. Freedman Presidential Scholars Program, the Dartmouth E.E. Just Program, the National Science Foundation under AAG award numbers 1211096 and 1515364, and by an Alfred P. Sloan Research Fellowship. R.C.H. acknowledges support from the National Science Foundation through grant number 1515404 and CAREER award 1554584, and from NASA through grants NNX16AN48G, NNX15AP24G, and NNX15AU32H.

This work makes use of data from the Wide-field Infrared Survey Explorer which is a joint project of the University of California, Los Angeles, and the Jet Propulsion Laboratory/California Institute of Technology, funded by the National Aeronautics and Space Administration.

This data makes use of data from the Sloan Digital Sky Survey. Funding for SDSS-III has been provided by the Alfred P. Sloan Foundation, the Participating Institutions, the National Science Foundation, and the U.S. Department of Energy Office of Science. The SDSS-III web site is http://www.sdss3.org/.

SDSS-III is managed by the Astrophysical Research Consortium for the Participating Institutions of the SDSS-III Collaboration including the University of Arizona, the Brazilian Participation Group, Brookhaven National Laboratory, Carnegie Mellon University, University of Florida, the French Participation Group, the German Participation Group, Harvard University, the Instituto de Astrofisica de Canarias, the Michigan State/Notre Dame/JINA Participation Group, Johns Hopkins University, Lawrence Berkeley National Laboratory, Max Planck Institute for Astrophysics, Max Planck Institute for Extraterrestrial Physics, New Mexico State University, New York University, Ohio State University, Pennsylvania State University, University of Portsmouth, Princeton University, the Spanish Participation Group, University of Tokyo, University of Utah, Vanderbilt University, University of Virginia, University of Washington, and Yale University.

This work makes use of data from United Kingdom InfraRed Telescope (UKIRT) Infrared Deep Sky Survey (UKIDSS). The UKIDSS project is defined in \citet{LawrenceUKIDSS}. UKIDSS uses the UKIRT Wide Field Camera \citep[WFCAM]{UKIRTWFC} and a photometric system described in \citet{UKIRTPhot}. The pipeline processing and science archive is described in \citet{WFCAM08}. 



\bibliographystyle{mnras}
\bibliography{research} 

\begin{thebibliography}{}
\makeatletter
\relax
\def\mn@urlcharsother{\let\do\@makeother \do\$\do\&\do\#\do\^\do\_\do\%\do\~}
\def\mn@doi{\begingroup\mn@urlcharsother \@ifnextchar [ {\mn@doi@}
  {\mn@doi@[]}}
\def\mn@doi@[#1]#2{\def\@tempa{#1}\ifx\@tempa\@empty \href
  {http://dx.doi.org/#2} {doi:#2}\else \href {http://dx.doi.org/#2} {#1}\fi
  \endgroup}
\def\mn@eprint#1#2{\mn@eprint@#1:#2::\@nil}
\def\mn@eprint@arXiv#1{\href {http://arxiv.org/abs/#1} {{\tt arXiv:#1}}}
\def\mn@eprint@dblp#1{\href {http://dblp.uni-trier.de/rec/bibtex/#1.xml}
  {dblp:#1}}
\def\mn@eprint@#1:#2:#3:#4\@nil{\def\@tempa {#1}\def\@tempb {#2}\def\@tempc
  {#3}\ifx \@tempc \@empty \let \@tempc \@tempb \let \@tempb \@tempa \fi \ifx
  \@tempb \@empty \def\@tempb {arXiv}\fi \@ifundefined
  {mn@eprint@\@tempb}{\@tempb:\@tempc}{\expandafter \expandafter \csname
  mn@eprint@\@tempb\endcsname \expandafter{\@tempc}}}

\bibitem[\protect\citeauthoryear{{Ahn} et~al.,}{{Ahn} et~al.}{2012}]{ahn12sdss}
{Ahn} C.~P.,  et~al., 2012, \mn@doi [\apjs] {10.1088/0067-0049/203/2/21}, \href
  {http://adsabs.harvard.edu/abs/2012ApJS..203...21A} {203, 21}

\bibitem[\protect\citeauthoryear{{Alam} et~al.,}{{Alam}
  et~al.}{2015}]{SDSSDR11+12}
{Alam} S.,  et~al., 2015, \mn@doi [\apjs] {10.1088/0067-0049/219/1/12}, \href
  {http://adsabs.harvard.edu/abs/2015ApJS..219...12A} {219, 12}

\bibitem[\protect\citeauthoryear{{Alexander} \& {Hickox}}{{Alexander} \&
  {Hickox}}{2012}]{alex12bh}
{Alexander} D.~M.,  {Hickox} R.~C.,  2012, \mn@doi [\nar]
  {10.1016/j.newar.2011.11.003}, \href
  {http://adsabs.harvard.edu/abs/2012NewAR..56...93A} {56, 93}

\bibitem[\protect\citeauthoryear{{Assef} et~al.,}{{Assef}
  et~al.}{2010}]{asse10agntemp}
{Assef} R.~J.,  et~al., 2010, \mn@doi [\apj] {10.1088/0004-637X/713/2/970},
  \href {http://adsabs.harvard.edu/abs/2010ApJ...713..970A} {713, 970}

\bibitem[\protect\citeauthoryear{{Baldwin}, {Phillips}  \&
  {Terlevich}}{{Baldwin} et~al.}{1981}]{bald81bpt}
{Baldwin} J.~A.,  {Phillips} M.~M.,   {Terlevich} R.,  1981, \mn@doi [\pasp]
  {10.1086/130766}, \href {http://adsabs.harvard.edu/abs/1981PASP...93....5B}
  {93, 5}

\bibitem[\protect\citeauthoryear{{Barret} et~al.,}{{Barret}
  et~al.}{2013}]{barret13athena}
{Barret} D.,  et~al., 2013, in {Cambresy} L.,  {Martins} F.,  {Nuss} E.,
  {Palacios} A.,  eds, SF2A-2013: Proceedings of the Annual meeting of the
  French Society of Astronomy and Astrophysics. pp 447--453 (\mn@eprint {arXiv}
  {1310.3814})

\bibitem[\protect\citeauthoryear{{Beck}, {Dobos}, {Budav{\'a}ri}, {Szalay}  \&
  {Csabai}}{{Beck} et~al.}{2016}]{Beck16drphot}
{Beck} R.,  {Dobos} L.,  {Budav{\'a}ri} T.,  {Szalay} A.~S.,   {Csabai} I.,
  2016, \mn@doi [\mnras] {10.1093/mnras/stw1009}, \href
  {http://adsabs.harvard.edu/abs/2016MNRAS.460.1371B} {460, 1371}

\bibitem[\protect\citeauthoryear{{Brandt} \& {Alexander}}{{Brandt} \&
  {Alexander}}{2010}]{bran10}
{Brandt} W.~N.,  {Alexander} D.~M.,  2010, \mn@doi [Proceedings of the National
  Academy of Science] {10.1073/pnas.0914151107}, \href
  {http://adsabs.harvard.edu/abs/2010PNAS..107.7184B} {107, 7184}

\bibitem[\protect\citeauthoryear{{Cardelli}, {Clayton}  \& {Mathis}}{{Cardelli}
  et~al.}{1989}]{Cardelli89Dust}
{Cardelli} J.~A.,  {Clayton} G.~C.,   {Mathis} J.~S.,  1989, \mn@doi [\apj]
  {10.1086/167900}, \href {http://adsabs.harvard.edu/abs/1989ApJ...345..245C}
  {345, 245}

\bibitem[\protect\citeauthoryear{{Carnall}}{{Carnall}}{2017}]{Carnall17SpectRes}
{Carnall} A.~C.,  2017, preprint, \href
  {http://adsabs.harvard.edu/abs/2017arXiv170505165C} {} (\mn@eprint {arXiv}
  {1705.05165})

\bibitem[\protect\citeauthoryear{{Casali} et~al.,}{{Casali}
  et~al.}{2007}]{UKIRTWFC}
{Casali} M.,  et~al., 2007, \mn@doi [\aap] {10.1051/0004-6361:20066514}, \href
  {http://adsabs.harvard.edu/abs/2007A%26A...467..777C} {467, 777}

\bibitem[\protect\citeauthoryear{{Collet} et~al.,}{{Collet}
  et~al.}{2016}]{Collet16feedback}
{Collet} C.,  et~al., 2016, \mn@doi [\aap] {10.1051/0004-6361/201526872}, \href
  {http://adsabs.harvard.edu/abs/2016A%26A...586A.152C} {586, A152}

\bibitem[\protect\citeauthoryear{{Cutri} \& {et al.}}{{Cutri} \& {et
  al.}}{2013}]{allWISE+13}
{Cutri} R.~M.,  {et al.} 2013, VizieR Online Data Catalog, \href
  {http://adsabs.harvard.edu/abs/2013yCat.2328....0C} {2328}

\bibitem[\protect\citeauthoryear{{DiPompeo}, {Bovy}, {Myers}  \&
  {Lang}}{{DiPompeo} et~al.}{2015}]{dipo15qsoz}
{DiPompeo} M.~A.,  {Bovy} J.,  {Myers} A.~D.,   {Lang} D.,  2015, \mnras\ in
  press (arXiv:1507.02884), \href
  {http://adsabs.harvard.edu/abs/2015arXiv150702884D} {}

\bibitem[\protect\citeauthoryear{{Eisenstein} et~al.,}{{Eisenstein}
  et~al.}{2011}]{sdssIII11}
{Eisenstein} D.~J.,  et~al., 2011, \mn@doi [\aj] {10.1088/0004-6256/142/3/72},
  \href {http://adsabs.harvard.edu/abs/2011AJ....142...72E} {142, 72}

\bibitem[\protect\citeauthoryear{{Fabian}}{{Fabian}}{2012}]{fabi12feed}
{Fabian} A.~C.,  2012, \mn@doi [\araa] {10.1146/annurev-astro-081811-125521},
  \href {http://adsabs.harvard.edu/abs/2012ARA%26A..50..455F} {50, 455}

\bibitem[\protect\citeauthoryear{{Galavis}, {Mendoza}  \& {Zeippen}}{{Galavis}
  et~al.}{1997}]{Galavis97}
{Galavis} M.~E.,  {Mendoza} C.,   {Zeippen} C.~J.,  1997, \mn@doi [\aaps]
  {10.1051/aas:1997344}, \href
  {http://adsabs.harvard.edu/abs/1997A%26AS..123..159G} {123}

\bibitem[\protect\citeauthoryear{{Gordon} \& {Clayton}}{{Gordon} \&
  {Clayton}}{1998}]{Gordon98SMC}
{Gordon} K.~D.,  {Clayton} G.~C.,  1998, \mn@doi [\apj] {10.1086/305774}, \href
  {http://adsabs.harvard.edu/abs/1998ApJ...500..816G} {500, 816}

\bibitem[\protect\citeauthoryear{{Haardt} \& {Maraschi}}{{Haardt} \&
  {Maraschi}}{1991}]{Haardt91Xray}
{Haardt} F.,  {Maraschi} L.,  1991, \mn@doi [\apjl] {10.1086/186171}, \href
  {http://adsabs.harvard.edu/abs/1991ApJ...380L..51H} {380, L51}

\bibitem[\protect\citeauthoryear{{Hainline}, {Hickox}, {Carroll}, {Myers},
  {DiPompeo}  \& {Trouille}}{{Hainline} et~al.}{2014}]{hain14salt}
{Hainline} K.~N.,  {Hickox} R.~C.,  {Carroll} C.~M.,  {Myers} A.~D.,
  {DiPompeo} M.~A.,   {Trouille} L.,  2014, \mn@doi [\apj]
  {10.1088/0004-637X/795/2/124}, \href
  {http://adsabs.harvard.edu/abs/2014ApJ...795..124H} {795, 124}

\bibitem[\protect\citeauthoryear{{Hambly} et~al.,}{{Hambly}
  et~al.}{2008}]{WFCAM08}
{Hambly} N.~C.,  et~al., 2008, \mn@doi [\mnras]
  {10.1111/j.1365-2966.2007.12700.x}, \href
  {http://adsabs.harvard.edu/abs/2008MNRAS.384..637H} {384, 637}

\bibitem[\protect\citeauthoryear{{Harrison} et~al.,}{{Harrison}
  et~al.}{2016}]{Harrison16KASHz}
{Harrison} C.~M.,  et~al., 2016, \mn@doi [\mnras] {10.1093/mnras/stv2727},
  \href {http://adsabs.harvard.edu/abs/2016MNRAS.456.1195H} {456, 1195}

\bibitem[\protect\citeauthoryear{{Hewett}, {Warren}, {Leggett}  \&
  {Hodgkin}}{{Hewett} et~al.}{2006}]{UKIRTPhot}
{Hewett} P.~C.,  {Warren} S.~J.,  {Leggett} S.~K.,   {Hodgkin} S.~T.,  2006,
  \mn@doi [\mnras] {10.1111/j.1365-2966.2005.09969.x}, \href
  {http://adsabs.harvard.edu/abs/2006MNRAS.367..454H} {367, 454}

\bibitem[\protect\citeauthoryear{{Hopkins}, {Hernquist}, {Cox}, {Di Matteo},
  {Robertson}  \& {Springel}}{{Hopkins} et~al.}{2006}]{hopk06apjs}
{Hopkins} P.~F.,  {Hernquist} L.,  {Cox} T.~J.,  {Di Matteo} T.,  {Robertson}
  B.,   {Springel} V.,  2006, \mn@doi [\apjs] {10.1086/499298}, \href
  {http://adsabs.harvard.edu/cgi-bin/nph-
  bib_query?bibcode=2006ApJS..163....1H&db_key=AST} {163, 1}

\bibitem[\protect\citeauthoryear{{Hopkins}, {Hernquist}, {Cox}  \& {Kere{\v
  s}}}{{Hopkins} et~al.}{2008}]{hopk08frame1}
{Hopkins} P.~F.,  {Hernquist} L.,  {Cox} T.~J.,   {Kere{\v s}} D.,  2008,
  \mn@doi [\apjs] {10.1086/524362}, \href
  {http://adsabs.harvard.edu/abs/2008ApJS..175..356H} {175, 356}

\bibitem[\protect\citeauthoryear{{Kauffmann} \& {Haehnelt}}{{Kauffmann} \&
  {Haehnelt}}{2000}]{kauf00merge}
{Kauffmann} G.,  {Haehnelt} M.,  2000, \mnras, \href
  {http://adsabs.harvard.edu/abs/2000MNRAS.311..576K} {311, 576}

\bibitem[\protect\citeauthoryear{{Kewley}, {Groves}, {Kauffmann}  \&
  {Heckman}}{{Kewley} et~al.}{2006}]{kewl06agn}
{Kewley} L.~J.,  {Groves} B.,  {Kauffmann} G.,   {Heckman} T.,  2006, \mn@doi
  [\mnras] {10.1111/j.1365-2966.2006.10859.x}, \href
  {http://adsabs.harvard.edu/abs/2006MNRAS.372..961K} {372, 961}

\bibitem[\protect\citeauthoryear{{Kobulnicky}, {Nordsieck}, {Burgh}, {Smith},
  {Percival}, {Williams}  \& {O'Donoghue}}{{Kobulnicky}
  et~al.}{2003}]{Kobulnicky03RSS}
{Kobulnicky} H.~A.,  {Nordsieck} K.~H.,  {Burgh} E.~B.,  {Smith} M.~P.,
  {Percival} J.~W.,  {Williams} T.~B.,   {O'Donoghue} D.,  2003, in {Iye} M.,
  {Moorwood} A.~F.~M.,  eds,  \procspie Vol. 4841, Instrument Design and
  Performance for Optical/Infrared Ground-based Telescopes. pp 1634--1644,
  \mn@doi{10.1117/12.460315}

\bibitem[\protect\citeauthoryear{{Komatsu} et~al.,}{{Komatsu}
  et~al.}{2011}]{komatsu2011}
{Komatsu} E.,  et~al., 2011, \mn@doi [\apjs] {10.1088/0067-0049/192/2/18},
  \href {http://adsabs.harvard.edu/abs/2011ApJS..192...18K} {192, 18}

\bibitem[\protect\citeauthoryear{{Lacy} et~al.,}{{Lacy}
  et~al.}{2013}]{lacy13spec}
{Lacy} M.,  et~al., 2013, \mn@doi [\apjs] {10.1088/0067-0049/208/2/24}, \href
  {http://adsabs.harvard.edu/abs/2013ApJS..208...24L} {208, 24}

\bibitem[\protect\citeauthoryear{{Lawrence} et~al.,}{{Lawrence}
  et~al.}{2007}]{LawrenceUKIDSS}
{Lawrence} A.,  et~al., 2007, \mn@doi [\mnras]
  {10.1111/j.1365-2966.2007.12040.x}, \href
  {http://adsabs.harvard.edu/abs/2007MNRAS.379.1599L} {379, 1599}

\bibitem[\protect\citeauthoryear{{Liu}, {Zakamska}, {Greene}, {Nesvadba}  \&
  {Liu}}{{Liu} et~al.}{2013}]{Liu13feedback}
{Liu} G.,  {Zakamska} N.~L.,  {Greene} J.~E.,  {Nesvadba} N.~P.~H.,   {Liu} X.,
   2013, \mn@doi [\mnras] {10.1093/mnras/stt1755}, \href
  {http://adsabs.harvard.edu/abs/2013MNRAS.436.2576L} {436, 2576}

\bibitem[\protect\citeauthoryear{{Mateos} et~al.,}{{Mateos}
  et~al.}{2012}]{mate12xmmwise}
{Mateos} S.,  et~al., 2012, \mnras\ in press (arXiv:1208.2530), \href
  {http://adsabs.harvard.edu/abs/2012arXiv1208.2530M} {}

\bibitem[\protect\citeauthoryear{{Mullaney}, {Alexander}, {Fine}, {Goulding},
  {Harrison}  \& {Hickox}}{{Mullaney} et~al.}{2013}]{Mullaney13Kinem}
{Mullaney} J.~R.,  {Alexander} D.~M.,  {Fine} S.,  {Goulding} A.~D.,
  {Harrison} C.~M.,   {Hickox} R.~C.,  2013, \mn@doi [\mnras]
  {10.1093/mnras/stt751}, \href
  {http://adsabs.harvard.edu/abs/2013MNRAS.433..622M} {433, 622}

\bibitem[\protect\citeauthoryear{{Nenkova}, {Sirocky}, {Ivezi{\'c}}  \&
  {Elitzur}}{{Nenkova} et~al.}{2008}]{Nenkova08Tor}
{Nenkova} M.,  {Sirocky} M.~M.,  {Ivezi{\'c}} {\v Z}.,   {Elitzur} M.,  2008,
  \mn@doi [\apj] {10.1086/590482}, \href
  {http://adsabs.harvard.edu/abs/2008ApJ...685..147N} {685, 147}

\bibitem[\protect\citeauthoryear{{Padovani et al.}}{{Padovani et
  al.}}{2017}]{PadovaniSub}
{Padovani et al.} 2017, \mn@doi [\aapr ~Submitted] {10.1093/mnras/stt751},
  \href {http://adsabs.harvard.edu/abs/2013MNRAS.433..622M} {}

\bibitem[\protect\citeauthoryear{{Reyes} et~al.,}{{Reyes}
  et~al.}{2008}]{reye08qso2}
{Reyes} R.,  et~al., 2008, \mn@doi [\aj] {10.1088/0004-6256/136/6/2373}, \href
  {http://adsabs.harvard.edu/abs/2008AJ....136.2373R} {136, 2373}

\bibitem[\protect\citeauthoryear{{Sanders} \& {Mirabel}}{{Sanders} \&
  {Mirabel}}{1996}]{sand96}
{Sanders} D.~B.,  {Mirabel} I.~F.,  1996, \mn@doi [\araa]
  {10.1146/annurev.astro.34.1.749}, \href
  {http://adsabs.harvard.edu/cgi-bin/nph-bib_query?bibcode=1996ARA%26A..34..749S&db_key=AST}
  {34, 749}

\bibitem[\protect\citeauthoryear{{Schmidt}}{{Schmidt}}{1963}]{schmidt63qso}
{Schmidt} M.,  1963, \mn@doi [\nat] {10.1038/1971040a0}, \href
  {http://adsabs.harvard.edu/abs/1963Natur.197.1040S} {197, 1040}

\bibitem[\protect\citeauthoryear{{Secrest}, {Dudik}, {Dorland}, {Zacharias},
  {Makarov}, {Fey}, {Frouard}  \& {Finch}}{{Secrest}
  et~al.}{2015}]{Secrest15WISE}
{Secrest} N.~J.,  {Dudik} R.~P.,  {Dorland} B.~N.,  {Zacharias} N.,  {Makarov}
  V.,  {Fey} A.,  {Frouard} J.,   {Finch} C.,  2015, \mn@doi [\apjs]
  {10.1088/0067-0049/221/1/12}, \href
  {http://adsabs.harvard.edu/abs/2015ApJS..221...12S} {221, 12}

\bibitem[\protect\citeauthoryear{{Smith}, {Nordsieck}, {Burgh}, {Percival},
  {Williams}, {O'Donohue}, {O'Connor}  \& {Schier}}{{Smith}
  et~al.}{2006}]{Smith06RSS}
{Smith} M.~P.,  {Nordsieck} K.~H.,  {Burgh} E.~B.,  {Percival} J.~W.,
  {Williams} T.~B.,  {O'Donohue} D.,  {O'Connor} J.,   {Schier} J.~A.,  2006,
  in Society of Photo-Optical Instrumentation Engineers (SPIE) Conference
  Series. p. 62692A, \mn@doi{10.1117/12.672415}

\bibitem[\protect\citeauthoryear{{Stern} et~al.,}{{Stern}
  et~al.}{2012}]{ster12wise}
{Stern} D.,  et~al., 2012, \mn@doi [\apj] {10.1088/0004-637X/753/1/30}, \href
  {http://adsabs.harvard.edu/abs/2012ApJ...753...30S} {753, 30}

\bibitem[\protect\citeauthoryear{{Sun}, {Greene}  \& {Zakamska}}{{Sun}
  et~al.}{2017}]{Sun17Kinem}
{Sun} A.-L.,  {Greene} J.~E.,   {Zakamska} N.~L.,  2017, \mn@doi [\apj]
  {10.3847/1538-4357/835/2/222}, \href
  {http://adsabs.harvard.edu/abs/2017ApJ...835..222S} {835, 222}

\bibitem[\protect\citeauthoryear{{Trouille}, {Barger}  \&
  {Tremonti}}{{Trouille} et~al.}{2011}]{trou11optx}
{Trouille} L.,  {Barger} A.~J.,   {Tremonti} C.,  2011, \mn@doi [\apj]
  {10.1088/0004-637X/742/1/46}, \href
  {http://adsabs.harvard.edu/abs/2011ApJ...742...46T} {742, 46}

\bibitem[\protect\citeauthoryear{{Urry} \& {Padovani}}{{Urry} \&
  {Padovani}}{1995}]{urry95}
{Urry} C.~M.,  {Padovani} P.,  1995, \pasp, \href
  {http://adsabs.harvard.edu/cgi-bin/nph-bib_query?bibcode=1995PASP..107..803U&db_key=AST}
  {107, 803}

\bibitem[\protect\citeauthoryear{{Vega Beltr{\'a}n}, {Pizzella}, {Corsini},
  {Funes}, {Zeilinger}, {Beckman}  \& {Bertola}}{{Vega Beltr{\'a}n}
  et~al.}{2001}]{Vega01Kinem}
{Vega Beltr{\'a}n} J.~C.,  {Pizzella} A.,  {Corsini} E.~M.,  {Funes} J.~G.,
  {Zeilinger} W.~W.,  {Beckman} J.~E.,   {Bertola} F.,  2001, \mn@doi [\aap]
  {10.1051/0004-6361:20010625}, \href
  {http://adsabs.harvard.edu/abs/2001A%26A...374..394V} {374, 394}

\bibitem[\protect\citeauthoryear{{Vignali}}{{Vignali}}{2014}]{Vignali14obsc}
{Vignali} C.,  2014, in {Mickaelian} A.~M.,  {Sanders} D.~B.,  eds,  IAU
  Symposium Vol. 304, Multiwavelength AGN Surveys and Studies. pp 132--138
  (\mn@eprint {arXiv} {1401.5061}), \mn@doi{10.1017/S1743921314003548}

\bibitem[\protect\citeauthoryear{{Weisskopf}, {Gaskin}, {Tananbaum}  \&
  {Vikhlinin}}{{Weisskopf} et~al.}{2015}]{Weisskopf-Surveyor}
{Weisskopf} M.~C.,  {Gaskin} J.,  {Tananbaum} H.,   {Vikhlinin} A.,  2015, in
  EUV and X-ray Optics: Synergy between Laboratory and Space IV.  (\mn@eprint
  {arXiv} {1505.00814})

\bibitem[\protect\citeauthoryear{{Wright} et~al.,}{{Wright}
  et~al.}{2010}]{wrig10wise}
{Wright} E.~L.,  et~al., 2010, \mn@doi [\aj] {10.1088/0004-6256/140/6/1868},
  \href {http://adsabs.harvard.edu/abs/2010AJ....140.1868W} {140, 1868}

\bibitem[\protect\citeauthoryear{{Yuan}, {Strauss}  \& {Zakamska}}{{Yuan}
  et~al.}{2016}]{Yuan16T2QSO}
{Yuan} S.,  {Strauss} M.~A.,   {Zakamska} N.~L.,  2016, \mn@doi [\mnras]
  {10.1093/mnras/stw1747}, \href
  {http://adsabs.harvard.edu/abs/2016MNRAS.462.1603Y} {462, 1603}

\bibitem[\protect\citeauthoryear{{Zakamska} et~al.}{{Zakamska}
  et~al.}{2003}]{zakamskaetal2003}
{Zakamska} N.~L.,  et~al., 2003, \mn@doi [\aj] {10.1086/378610}, \href
  {http://adsabs.harvard.edu/abs/2003AJ....126.2125Z} {126, 2125}

\makeatother
\end{thebibliography}

\bsp	
\label{lastpage}
\end{document}